\documentclass[iop]{emulateapj}%
\usepackage{amsmath}%
\usepackage{amsfonts}%
\usepackage{amssymb}%
\usepackage{graphicx}%
\usepackage{epsfig}%
\usepackage{caption}
\usepackage{natbib}

\begin{document}

\slugcomment{\em Accepted for Publication in the Astrophysical Journal}

\title{The Dust and Molecular Gas in the Brightest Cluster Galaxy in MACS 1931.8-2635}

\author{Kevin Fogarty\altaffilmark{1}\altaffilmark{2}}
\author{Marc Postman\altaffilmark{2}}
\author{Yuan Li\altaffilmark{3}}
\author{Helmut Dannerbauer\altaffilmark{4}\altaffilmark{5}}
\author{Hauyu Baobab Liu\altaffilmark{6}}
\author{Megan Donahue\altaffilmark{7}}
\author{Bodo Ziegler\altaffilmark{8}}
\author{Anton Koekemoer\altaffilmark{2}}
\author{Brenda Frye\altaffilmark{9}}

\altaffiltext{1}{Division of Physics, Math, and Astronomy, California Institute of Technology, Pasadena, CA, USA}
\altaffiltext{2}{Space Telescope Science Institute, Baltimore, MD, USA}
\altaffiltext{3}{Department of Astronomy, University of California, Berkeley, CA, USA}
\altaffiltext{4}{Instituto de Astrofisica de Canarias (IAC), E-38205 La Laguna, Tenerife, Spain}
\altaffiltext{5}{Universidad de La Laguna, Dpto. Astrofisica, E-38206 La Laguna, Tenerife, Spain}
\altaffiltext{6}{Academia Sinica, Institute of Astronomy and Astrophysics, Taiwan}
\altaffiltext{7}{Department of Physics and Astronomy, Michigan State University, East Lansing, MI, USA}
\altaffiltext{8}{Department of Astrophysics, University of Vienna, Austria}
\altaffiltext{9}{Department of Astronomy and Steward Obs., University of Arizona, Tucson, AZ, USA}

\begin{abstract}
We present new ALMA observations of the molecular gas and far-infrared continuum around the brightest cluster galaxy (BCG) in the cool-core cluster MACS 1931.8-2635. Our observations reveal $1.9 \pm 0.3 \times 10^{10}$ M$_{\odot}$ of molecular gas, on par with the largest known reservoirs of cold gas in a cluster core. We detect CO(1-0), CO(3-2), and CO(4-3) emission from both diffuse and compact molecular gas components that extend from the BCG center out to $\sim30$ kpc to the northwest, tracing the UV knots and H$\alpha$ filaments observed by \textit{HST}. Due to the lack of morphological symmetry, we hypothesize that the $\sim300$ km\,s$^{-1}$ velocity of the CO in the tail is not due to concurrent uplift by AGN jets, rather we may be observing the aftermath of a recent AGN outburst. The CO spectral line energy distribution suggests that molecular gas excitation is influenced by processes related to both star formation and recent AGN feedback. Continuum emission in Bands 6 and 7 arises from dust and is spatially coincident with young stars and nebular emission observed in the UV and optical. We constrain the temperature of several dust clumps to be $\lesssim 10$ K, which is too cold to be directly interacting with the surrounding $\sim 4.8$ keV intracluster medium (ICM). The cold dust population extends beyond the observed CO emission and must either be protected from interacting with the ICM or be surrounded by local volumes of ICM that are several keV colder than observed by \textit{Chandra}.
\end{abstract}


\section{Introduction}

Brightest cluster galaxies (BCGs) in cool-core clusters of galaxies have been observed harboring reservoirs of cold molecular gas with masses up to several $10^{10}$ M$_{\odot}$ \citep{2001Edge_MolecularGas, 2003Salome_MolecularGas, 2003Edge_MolecularGas, 2008ODea_BCGSF}. These gas reservoirs have been reliably detected using CO collisional lines as a tracer of the predominantly H$_{2}$ gas, and exhibit a wide range of morphologies and dynamical characteristics. In some cool-core BCGs (such as Abell 1835), this gas appears to be clearly associated with AGN jets and star formation, while in others (such as NGC 1275), the gas is distributed in narrow filaments that are not currently forming stars \citep{2008Salome_NGC1275, 2014McNamara_A1835, 2017Lim_Perseus}.

The BCGs that possess these gas reservoirs often exhibit highly extended star formation and radio jets from active galactic nuclei (AGN) that excavate large ($>10$ kpc-wide) cavities in the surrounding intracluster medium (ICM) \citep{2006Rafferty_Feedback, 2007McNamara_AGNFeedback,  2012Fabian_AGNFeedback}. These features are thought to arise from AGN feedback-regulated cooling in the ICM, whereby some ICM condensation occurs but energy injected by the AGN into the ICM offsets radiative cooling and prevents the formation of the 100s-1000s M$_{\odot}$ yr$^{-1}$ cooling flows that were originally predicted to occur in these systems \citep[e.g.][]{1994Fabian_CoolingFlow, 2000McNamara_ChandraHydraA, 2003Peterson_CoolingFlow, 2006Peterson_CoolingFlow, 2012McNamara_MechanicalFeedback}.

Extensive modelling of mechanical-mode AGN feedback in cool core clusters reveals that while energy deposited in the ICM by AGN jets balances radiative cooling in a global, time averaged sense, locally the jets can stimulate gas condensation by pushing low-entropy volumes of ICM into regions of substantially higher ambient entropy \citep{2014Li_ColdClumps, 2015Gaspari_AGNFeedback, 2015Prasad_Feedback, 2016Yang_Feedback, 2016Meece_AGNFeedback, 2016Voit_Model, 2017Li_AGNHeating, 2018Gaspari_CondensationTracers, 2018Voit_Turbulence}. The resulting atomic and molecular gas condensate then precipitates onto the galactic nucleus, fueling starbursts and further AGN activity \citep{2013Gaspari_CCA, 2014Li_ColdClumps, 2015Li_SFAGN, 2016Gaspari_Condensation, 2016Voit_Model}. UV, optical, and near-IR observations of cool-core BCGs reveal star formation rates (SFRs) ranging from $\sim 1$ to several $100$ M$_{\odot}$ yr$^{-1}$, while mid-IR observations of the rovibrational states of H$_{2}$ and sub-mm observations of CO confirm the presence of molecular gas reservoirs as large as $\sim 10^{10}$ M$_{\odot}$ \citep{1989McNamara_CCSFR, 2008ODea_BCGSF, 2011Donahue_H2, 2015Tremblay_BCGUV, 2015Donahue_CLASH, 2016Loubser_BCGSF, 2018McDonald_CoolingFlow}.

Recent observations of BCGs with the Atacama Large Millimeter Array (ALMA) have found evidence of CO molecular gas tracing AGN jet outflows, providing observational support for the notion that molecular gas reservoirs in these systems are fueled by $> 10^{7}$ K gas that condenses as it is being uplifted by the jets \citep{2014McNamara_A1835, 2014Russell_A1664, 2016Russell_Phoenix, 2017Russell_A1795}. CO has also been observed in absorption along the line of sight to the AGN of BCGs, providing direct evidence for the accretion of cold gas onto the AGN \citep{2016Tremblay_Accretion, 2018Tremblay_ColdGas, 2019Rose_HydraA}. 

While the molecular gas reservoirs in many of the BCGs observed by ALMA appear to be consistent with forming out of the residual condensation in systems where AGN feedback largely balances cooling, a handful of systems show evidence of more extreme cooling and star formation. In particular, the BCG in the Phoenix cluster ($z = 0.597$) is host to a massive starburst event (SFR $\sim 450-2700$ M$_{\odot}$ yr$^{-1}$) on par with the predicted ICM radiative cooling rate \citep{2014McDonald_PhoenixGas, 2017Mittal_Phoenix}. This starburst may be short-lived compared to star formation in most other cool-core BCGs. The depletion timescales of the molecular gas in the majority of cool-core BCGs are on the order of a Gyr, suggesting these systems are capable of sustaining continuous modest star formation \citep{2008ODea_BCGSF}. However, in the Phoenix cluster the gas depletion timescale is $\lesssim 30$ Myr \citep{2014McDonald_PhoenixGas}. `Extreme' cool-core BCGs appear to be more common at higher redshifts ($z \gtrsim 0.6$), when BCGs tended to form stars at higher rates and, like the Phoenix BCG, tend to host an X-ray loud AGN \citep{2013HL_AGN, 2016McDonald_Evolution, 2018Cooke_Evolution}.

We can improve our understanding of the formation and evolution of gas and dust in and around cool-core BCGs by observing constituents of the potentially cooling gas (e.g., atomic gas, molecular gas, dust, and ICM plasma) in systems with a variety of feedback intensities and across various stages of cluster evolution. Such observations will also reveal how the surrounding gas and dust interact with the AGN and how that interaction affects the balance between radiative cooling and feedback in cool-core clusters. 

Extensive multiwavelength observations for a modest-sized sample of cool-core BCGs have been acquired for the Cluster Lensing And Supernova survey with Hubble (CLASH) program \citep{2012Postman_CLASH}. We have used these data to identify the cluster MACS 1931.8-2635 as an interesting candidate in which to study molecular gas and dust during an episode of elevated feedback and star formation in a cool core. We obtained ALMA observations of this cluster to further investigate the process of dust and molecular gas formation in BCGs and characterize the spatial distribution, energetics, and kinematics of the dust and molecular gas, and their relationship to the ICM, during episodes of intense starburst activity and AGN feedback.

The BCG in the CLASH cluster in MACS 1931.8-2635 (hereafter MACS 1931) has a SFR of $\sim 250$ M$_{\odot}$ yr$^{-1}$ and an X-ray loud AGN, suggesting it may be an example of `extreme' BCG condensation like the Phoenix \citep{2015Santos_AGN, 2017Fogarty_CLASH}. However, at a redshift of $z = 0.3525$, MACS 1931 is more nearby than typical quasar-mode BCGs with large SFRs. If MACS 1931 is part of the observed trend in BCG feedback evolution, it may be transitioning from the X-ray loud quasar-mode cooling and feedback characteristic of higher redshift cool core clusters to the X-ray quiet feedback mode typical of lower redshift cool cores. By studying MACS 1931, we hope to gain a better handle on how cooling and feedback in clusters progresses over time.

We present our investigation of the molecular gas and dust in the BCG of MACS 1931 using ALMA observations in Bands 3, 6, and 7 obtained during cycles 4 and 5. We observe the (1-0), (3-2) and (4-3) transitions of CO, along with dust continuum emission at rest-frame frequencies of 336 GHz and 468.5 GHz. These observations enable us to obtain intensity maps of CO and dust and to examine the dynamical state and velocity structure of molecular gas in the BCG. By modelling the dust emission as a greybody, we are also able to place rudimentary constraints on the dust temperature.

In Section 2, we summarize our ALMA dataset and describe the sources of all archival data used in this paper. In Section 3, we document our data reduction procedures, and in Section 4 we present the results of our observations. In Section 5, we discuss implications for the formation of multiphase gas (i.e. gas consisting of intermingled atomic, molecular, and hot plasma components) and dust in MACS 1931, and the possible relationships between dust, molecular gas, and features observed in the optical, X-ray, and radio at 1.5 GHz. We summarize our findings in Section 6. We adopt a $\Lambda$CDM cosmology throughout, with $\Omega_{m} = 0.3$, $\Omega_{\Lambda} = 0.7$, ${\rm H}_{\rm 0} = 70$ km\,s$^{-1}$\,Mpc$^{-1}$, and $h = 0.7$. With these cosmological parameters, 1$''$ subtends 4.962 kpc at the spectroscopic redshift of the BCG in MACS 1931, $z = 0.3525$.

\section{Data}

\subsection{ALMA Observations}

\begin{table*}%[]  
\footnotesize  
\caption{ALMA Observations}
\label{table:Observations}  
\vspace{1mm}  
\centering  
{  
\begin{tabular}{c|c|c|c|c|c|c}  
     &      & Observation  & Integration & Spectral & Spatial & Max.  \\
Band & Array & Start Date        &     Time & Resolution & Resolution & Angular Scale \\
& & &  (sec) & (MHz) & (arcsec) & (arcsec) \\
\hline  
\hline  
3 & 12 m & 2017-03-13 11:19:43 & 3084.48 & 7.8 & 3.6 & 19 \\
3 & 12 m & 2017-07-23 06:53:26 & 10281.6 & 7.8 & 0.54 & 7.0 \\
\hline
6 & 7 m & 2017-10-20 00:03:27 & 1723.68 & 31.25 & 5.2 & 20 \\
6 & 12 m & 2018-04-08 09:50:15 & 725.76 & 31.25 & 0.74 & 4.0 \\
\hline
7 & 7 m & 2017-10-10 00:16:31 & 6804 & 31.25 & 4.2 & 15 \\
7 & 12 m & 2018-05-15 07:21:43 & 2721.6 & 31.25 & 0.81 & 3.8 \\
\end{tabular}  
\begin{flushleft}
\end{flushleft}  
}  
\end{table*}

Our ALMA Band 3 observations were obtained in cycle 4 (Project ID 2016.1.00784.S). Our ALMA Bands 6 and 7 observations, including Atacama Compact Array (ACA) observations, were obtained in cycle 5 (Project ID 2017.1.01205.S). Basic information about these observations are summarized in Table \ref{table:Observations}. 

The Band 3 observations combined two ALMA 12 m array configurations, which yielded a 15-3700 m baseline range and a 13366 s total integration time. The minimum possible synthesized beam and the   recoverable angular scale are $0\farcs54$ and 18.8$''$, respectively. We configured our Band 3 spectral windows (SPWs) to center on sky frequencies of 85.3 GHz, 87.0 GHz, 97.0 GHz, and 99.0 GHz. The spectral channel width is 3.9 MHz (i.e., spectral resolution is 7.8 MHz), which corresponds to $\sim$11.7 km\,s$^{-1}$ velocity width.

We carried out one set of ALMA 12 m array observations at Band 6, with a baseline range of 15-500 m. It was complemented with a set of ACA observations with an identical spectral setup. In these observations, the Band 6 SPWs were configured to be centered on sky frequencies of 239.5 GHz, 241.5 GHz, 255.5 GHz, and 257.5 GHz. The spectral channel width was 15.625 MHz ($\sim$18.1 km\,s$^{-1}$). The on-source time for the 12 m array and the ACA observations were 726 s and 1724 s, respectively. Combining these observations yielded a $0\farcs74$ minimum possible synthesized beam and a $20\farcs4$ largest recoverable angular scale.

Similarly, we obtained one set of ALMA 12 m array observations at Band 7, with a baseline range of 15-314 m, paired with the complementary ACA observations. We configured the Band 7 SPWs to center on sky frequencies of 339.5 GHz, 341.4 GHz, 351.5 GHz, and 353.5 GHz. The spectral channel width was 15.625 MHz ($\sim$13.2 km\,s$^{-1}$). The on-source time for the 12 m array and ACA observations were 2722 s and 6804 s, respectively. Combining these observations yielded a $0\farcs81$ minimum possible synthesized beam and a $14\farcs6$ largest recoverable angular scale.

\subsection{HST Data}

We use the 16 bands of \textit{HST} photometry for the MACS J1931.8-2635 BCG that were obtained between rest wavelengths of $1745 \textrm{\AA}$ and $1.14 \mu$m as part of the CLASH program. These observations are detailed in \cite{2012Postman_CLASH} and the reduced dataset is described in \cite{2015Fogarty_CLASH, 2017Fogarty_CLASH}.

\section{Methods}

\subsection{ALMA Datacube Reductions}\label{Sec:Dust_Methods}

Cycle 4 observations were reduced using the CASA software package version 4.7.2 (pipeline r39732); cycle 5 observations were reduced using CASA version 5.1.1-5 (pipeline version r40896) \citep{2007McMullin_CASA}. After regenerating the pipeline-calibrated measurement sets, we used the task {\tt concat} to concatenate the calibrated measurement sets at each band. We used the task {\tt plotms} to visually inspect the spectra of our target source, and finally, used the task {\tt uvcontsub} to generate the continuum data from the line-free spectral channels and to produce continuum-subtracted CO line data.

We independently created images with the Band 3, 6 and 7 line-free continuum data using the multi-frequency synthesis (MFS) mode of the {\tt clean} task.
The visibilities were naturally weighted, and the image pixel size was $0\farcs075$. We produced clean masks using {\tt clean} in interactive mode to identify regions with significant source emission.

Flux in the Band 3 continuum is primarily due to the AGN. Flux in the Band 6 and 7 continuum images is due to thermal emission from extended dust and synchrotron and hot dust emission from the AGN. The AGN in our ALMA data is point-like and is most prominent in Band 3. To yield the dust continuum images, we subtracted the point source in each band using the image arithmetic task {\tt fitcomponents} to fit a Gaussian model to the point source with a uniform background level. The full widths at half maximum along the major and minor axes and the position angle of the Gaussian were fixed based on the synthesized beam of each image. We fit the peak coordinates of the point source, along with the source flux and background. The coordinates of the point source were the same in both bands. The point source flux at Band 6 is $4.19\pm0.42$ mJy, roughly equal to the $4.28\pm 0.43$ mJy measured in Band 7, ruling out the possibility that the emission is primarily due to greybody emission by hot dust in the AGN torus or unresolved circumnuclear region.

We ran {\tt clean} in `velocity' mode on each continuum-subtracted line dataset, using natural weighting, a $0\farcs075$ pixel size, and a velocity channel width of 25 km\,s$^{-1}$. The achieved synthesized beams for Band 3, Band 6, and Band 7 are 0\farcs82$\times$0\farcs53,\  0\farcs87$\times$0\farcs72, and 0\farcs94$\times$0\farcs78. In cases where we needed to produce datacubes with identical angular resolution from different datasets, we selected a common synthesized beam that encompassed the natural restoring beam for each band and smoothed using {\tt imsmooth} with {\tt targetres} = True. To study the extended, faint molecular gas structures around MACS 1931 BCG, we also produced {\it uv}-tapered CO image cubes to yield better brightness temperature sensitivities.

All images or image cubes used in our scientific analysis have been primary beam corrected.

\subsection{Astrometric Alignment Accuracy}

Our analyses of the {\it HST} and ALMA data requires that the images be aligned to a common astrometric reference frame. The astrometric reference frames for both {\it HST} and ALMA are calibrated to the International Celestial Reference System (ICRS). To assess any systematic offsets in the sky positions of features in our field, we cross-matched our {\it HST} astrometry to entries from the Gaia DR2 catalog \citep{2018GAIA_DR2} and find a mean offset of 95 mas with an rms of 26 mas. The maximum astrometric error in our ALMA data is estimated to be approximately 5\% of the synthesized beam width, which corresponds to 28 mas, 37 mas, and 41 mas, respectively, for bands 3, 6, and 7. For our study of the BCG in MACS 1931, any cross-comparisons of aligned features in our {\it HST} and ALMA data are performed on angular scales larger than 200 mas. This ensures that any systematic astrometric alignment errors between the {\it HST} and ALMA datasets will not be a significant concern.

\section{Results}

\subsection{CO Line Emission}\label{sec:CO_Results}

\begin{figure*}
\begin{center}
\begin{tabular}{c}
\includegraphics[height=5.8cm]{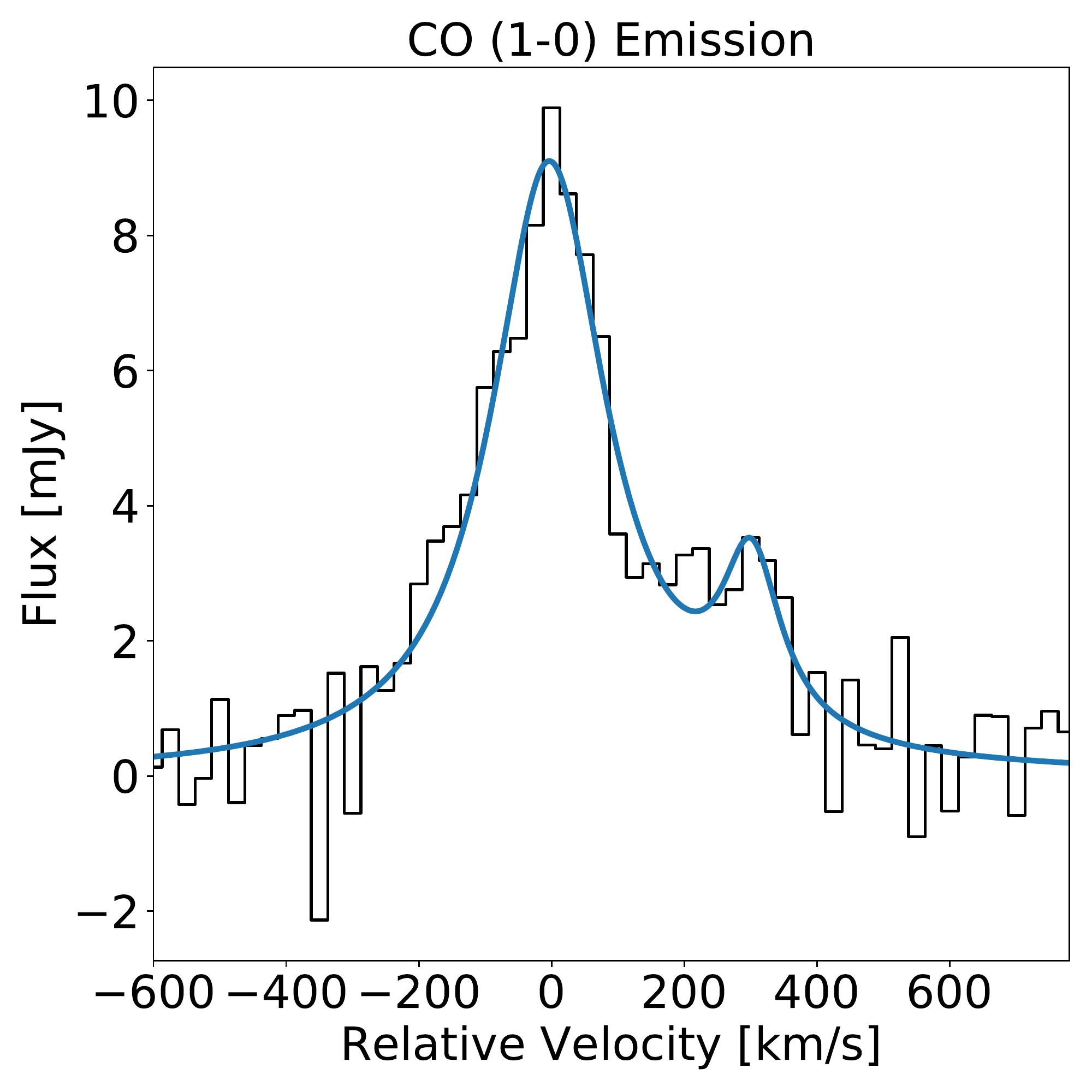}
\includegraphics[height=5.8cm]{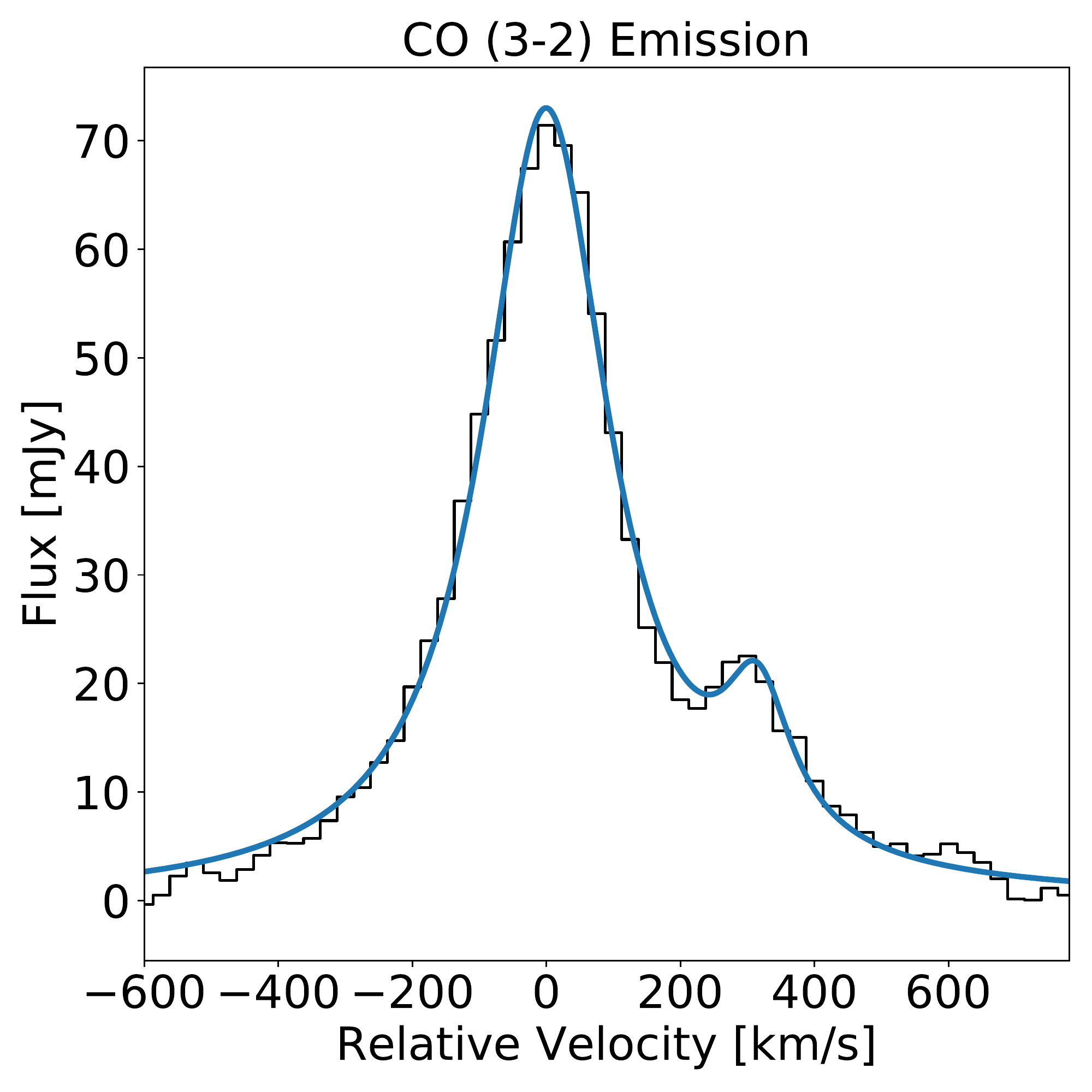}
\includegraphics[height=5.8cm]{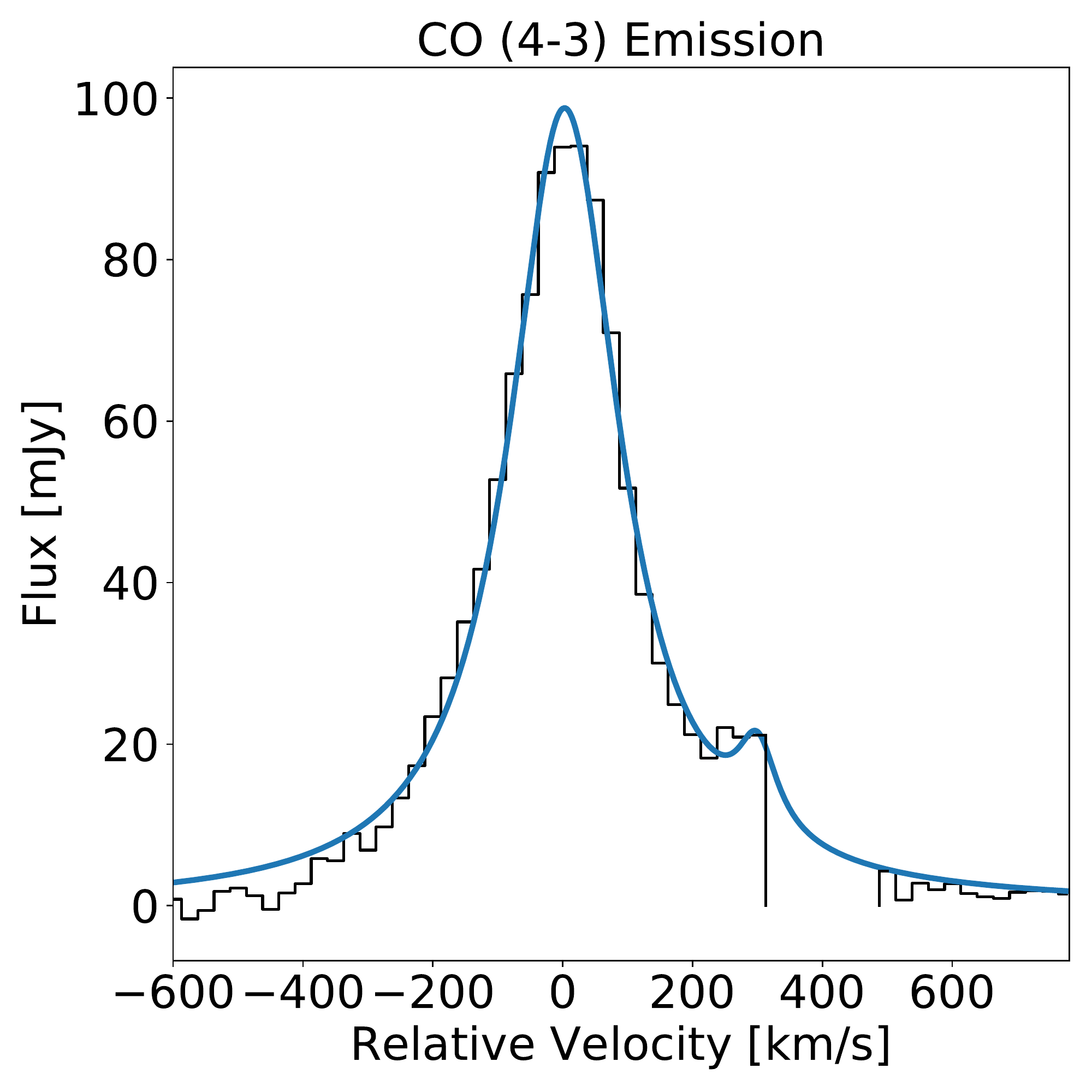}
\end{tabular}
\end{center}
\caption[]
{ \label{fig:CO_Lineplots} Emission line spectra for CO(1-0), CO(3-2), and CO(4-3) integrated across the MACS 1931 BCG. The observed spectra are shown in black and best-fit Lorentzian profiles are shown in blue. The parameters of the  best-fit two-component Lorentzian profiles are summarized in Table \ref{table:Line_Params}. The secondary peak in the CO(4-3) spectrum is partially obscured by the spectral window gap in Band 7.}
\end{figure*}

\begin{figure*}
\begin{center}
\begin{tabular}{c}
\includegraphics[height=18.6cm]{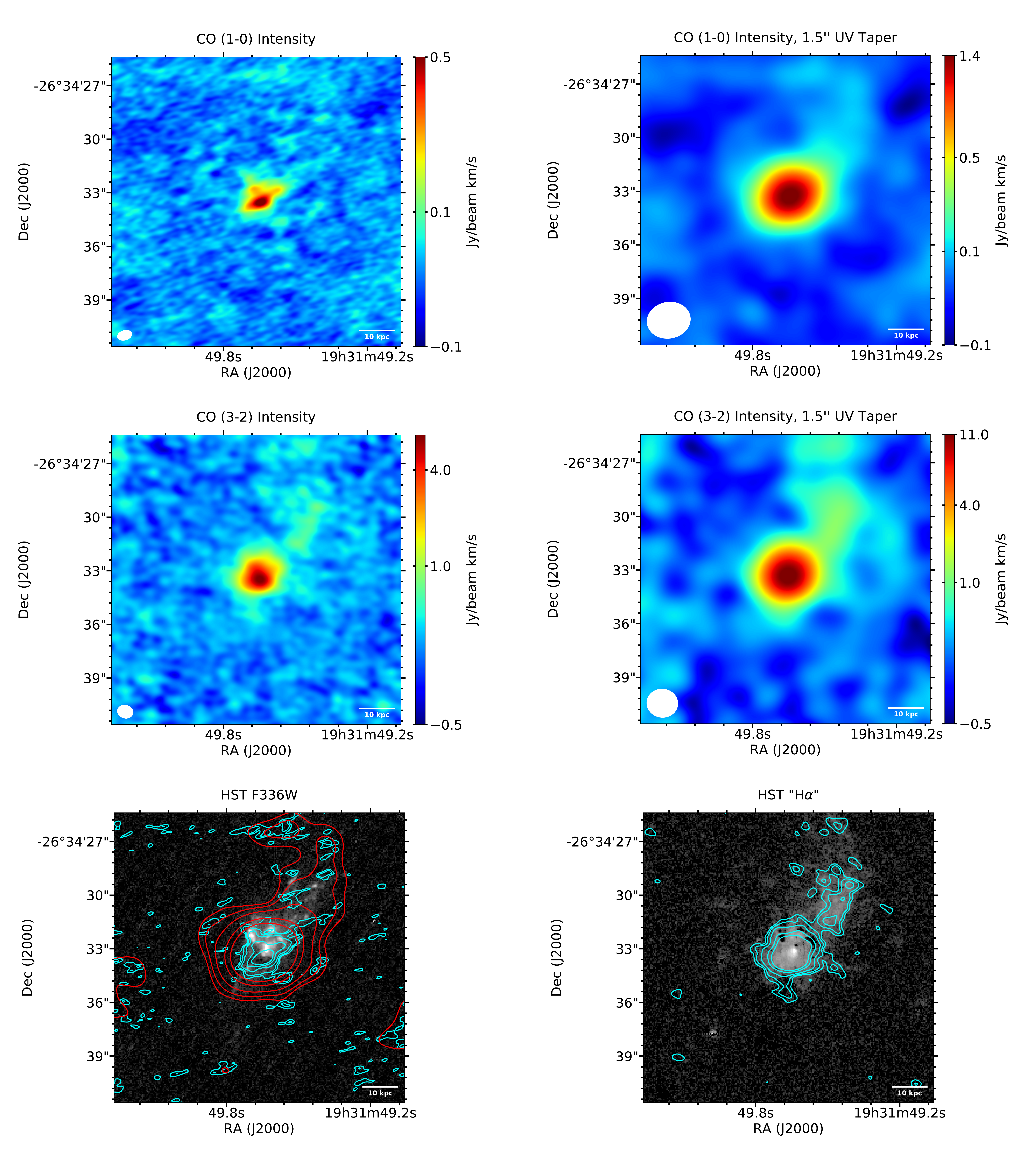}
\end{tabular}
\end{center}
\caption[]
{ \label{fig:CO_Images} \textit{Top Row:} CO(1-0) intensity maps with and without a $1\farcs5$ {\it uv}-taper applied to the cleaned image. The left image shows the intensity map with a naturally weighted clean synthesized beam, and the right image shows the map after the {\it uv}-taper is applied in order to reveal the extended emission to the northwest. The solid white ellipse in the lower left depicts the beam size. \textit{Middle Row:} CO(3-2) intensity maps with and without the {\it uv}-taper. \textit{Bottom Row:} The left-hand graphic shows the HST F336W (rest-frame 250 nm) image of the BCG, with ALMA contours overlaid. Blue contours represent the naturally weighted CO(1-0) intensity distribution. Red contours represent the {\it uv}-tapered CO(1-0) distribution.  The right-hand graphic shows the HST broadband estimated H$\alpha$ + [\ion{N}{2}] image, with contours from the CO(3-2) naturally weighted image overlaid. Both the blue and red contours trace 2, 3, 5, 10, and 15$\sigma$ significance.}
\end{figure*}

Our ALMA campaign obtained spatially resolved line emission from the $J=1$ to $0$ transition of CO(1-0) as well as CO(3-2) and partial coverage of CO(4-3). The spatially integrated line spectra and the velocity integrated intensity maps are shown in Figure \ref{fig:CO_Lineplots} and Figure \ref{fig:CO_Images}, respectively. We measured spectra in regions where the total intensity is detected at the $2\sigma$ level or greater. We also re-ran imaging for CO(1-0) and CO(3-2) using a $1\farcs5$ {\it uv}-taper, which down-weighted noisier, higher spatial frequency baselines. Using the original and the {\it uv}-tapered images enabled us to resolve the compact and luminous knots of CO emission in the core of the BCG and the extended and faint emission features along the H$\alpha$ ``tail'' extending to the northwest. Total fluxes in the {\it uv}-tapered images are consistent with the un-tapered images.

Spatially integrated flux from the emission lines were fit to Lorentzian profiles, shown as the solid blue lines in Figure \ref{fig:CO_Lineplots}. All three lines have two velocity components that are redshifted at $\sim0$\,km\,s$^{-1}$ and $\sim300$\,km\,s$^{-1}$ relative to the optical emission lines in the MACS 1931 BCG\footnote{The heliocentric redshift of the MACS 1931 BCG used as the optical systemic reference is $z = 0.35248 \pm 0.00004$ and is derived by us from 10 optical emission lines found in a VLT/MUSE spectrum extracted in a $3\farcs0$ radius aperture centered on the BCG. The wavelength-calibrated 1-D spectrum of the BCG was provided to us by P. Rosati and M. Verdugo.}. As shown below, the more redshifted component is due in part to a fraction of the molecular gas flowing either inward or outward from the galaxy center, and in part to motions of gas clumps in the center of the galaxy. Based on the best model fit, the CO(1-0) line has an integrated flux of $3.45 \pm 0.47$ Jy\,km\,s$^{-1}$, the CO(3-2) line has a flux of $29.0 \pm 2.8$ Jy\,km\,s$^{-1}$, and the CO(4-3) line has a flux of $33.7 \pm 3.5$ Jy\,km\,s$^{-1}$. Best-fit parameters for the lines are summarized in Table \ref{table:Line_Params}. 

\begin{table*}%[]  
\footnotesize  
\caption{Integrated Emission Line Properties}
\label{table:Line_Params}  
\vspace{1mm}  
\centering  
{  
\begin{tabular}{l|c|c|c|c|c|c}%{lrrrrrr}  
               & Component 1     &             & C1    & Component 2     &             & C2\\
Emission Line & (C1) Flux$^{a}$  & C1 Velocity & FWHM  & (C2) Flux$^{a}$ & C2 Velocity & FWHM \\
  & (Jy km s$^{-1}$) & (km s$^{-1}$) & (km s$^{-1}$) & (Jy km s$^{-1}$) & (km s$^{-1}$) & (km s$^{-1}$) \\ 
\hline  
\hline  
CO(1-0) & $3.02 \pm 0.42$ & $-3.7 \pm 5.6$ & $212 \pm 18$ & $0.43 \pm 0.2$ & $301 \pm 14$ & $108 \pm 44$ \\
CO(3-2) & $26.4 \pm 2.7$ & $-0.7 \pm 1.5$ & $231.4 \pm 4.8$ & $2.58 \pm 0.51$ & $315.1 \pm 5.8$ & $123 \pm 19$ \\
CO(4-3)$^{b}$ & $32.2 \pm 3.4$ & $2.7 \pm 1.7$ & $207.9 \pm 5.2$ & $1.49 \pm 0.78$ & 310 & $82 \pm 37$ 
\end{tabular}  
\begin{flushleft}
$^{a}$ Uncertainties in integrated fluxes include a 10\% absolute calibration accuracy term. \\
$^{b}$ Parameters for component C2 for the CO(4-3) line are based on the best fit to the partially covered spectral line component with a model mean velocity fixed at 310 km\,s$^{-1}$. 
\end{flushleft}  
}  
\end{table*}

Compact knots in the naturally weighted CO(1-0) intensity image trace UV knots present in \textit{HST} WFC3 F225W-F390W photometry (see Figure \ref{fig:CO_Images}). These knots are also visible in the naturally weighted CO(3-2) image but are less apparent given the shorter baselines and consequently lower spatial resolution of the image. The faint, extended component of the CO(1-0) and CO(3-2) emission traces the distribution of H$\alpha$-\ion{N}{2} emission inferred by the broadband difference image from \cite{2015Fogarty_CLASH}. This faint component forms both a diffuse nebula around the dense knots and a ``tail" extending $\sim 30$ kpc to the northwest of the optical brightness peak of the BCG (hereafter referred to as the BCG core). To within the angular resolution of our CO imaging, the \textit{HST} and ALMA data demonstrate a spatial correlation between the atomic and molecular phases of gas in the BCG, and a correlation between dense knots of molecular gas and sites of star formation, suggesting star formation in the BCG is fueled by multiphase material.

\begin{figure*}
\begin{center}
\begin{tabular}{c}
\includegraphics[height=6.5cm]{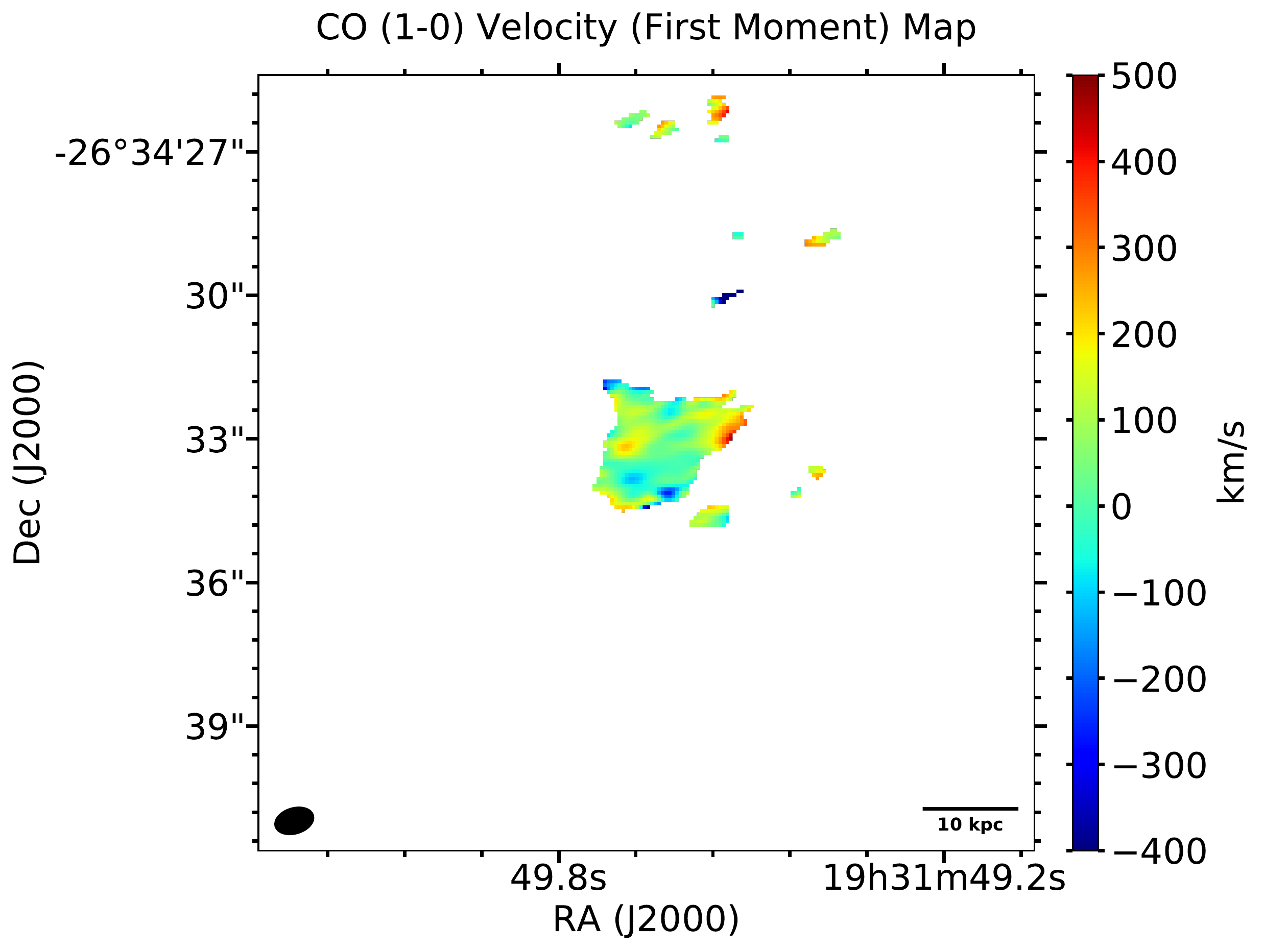}
\includegraphics[height=6.5cm]{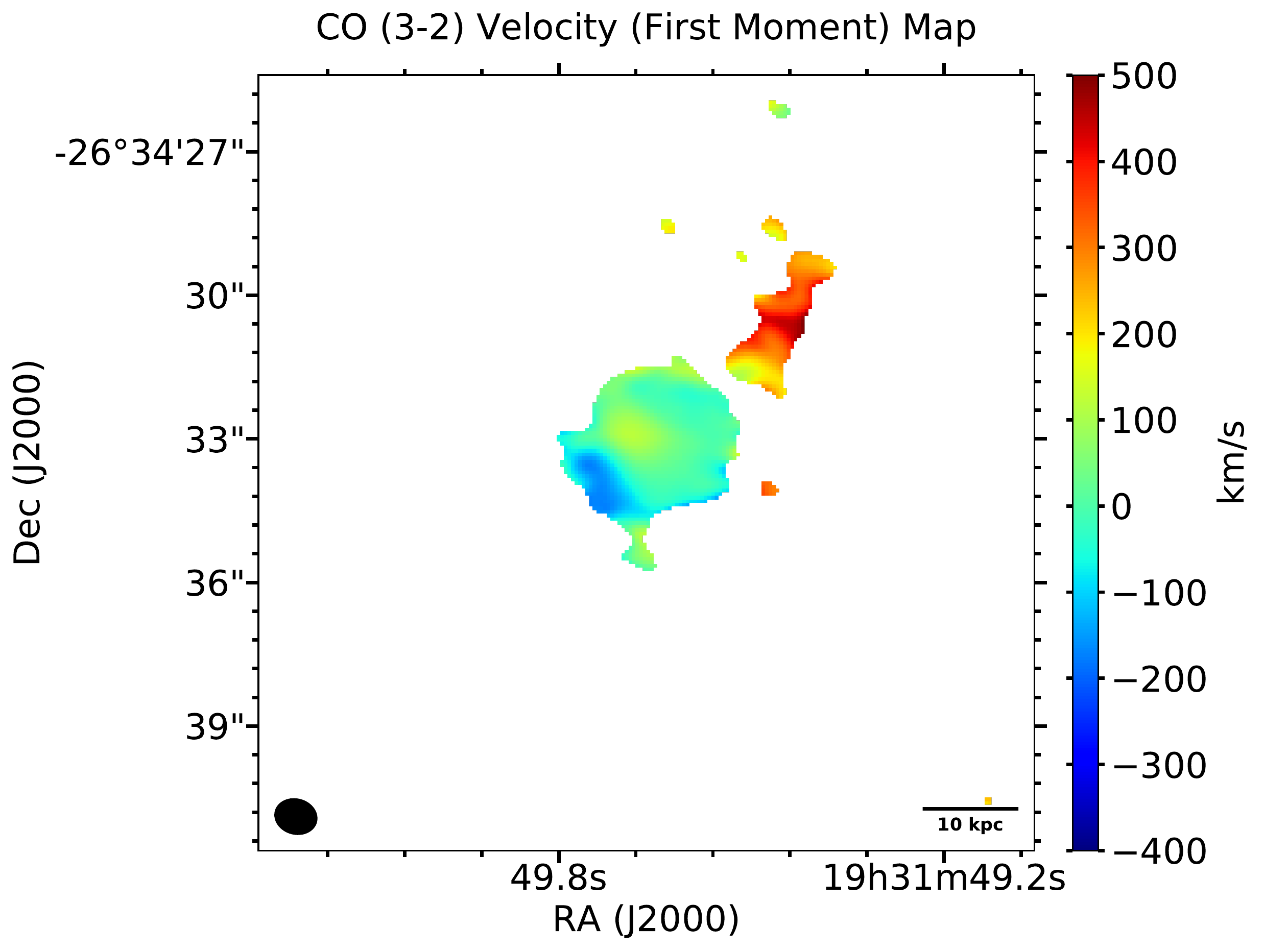}
\\
\includegraphics[height=6.5cm]{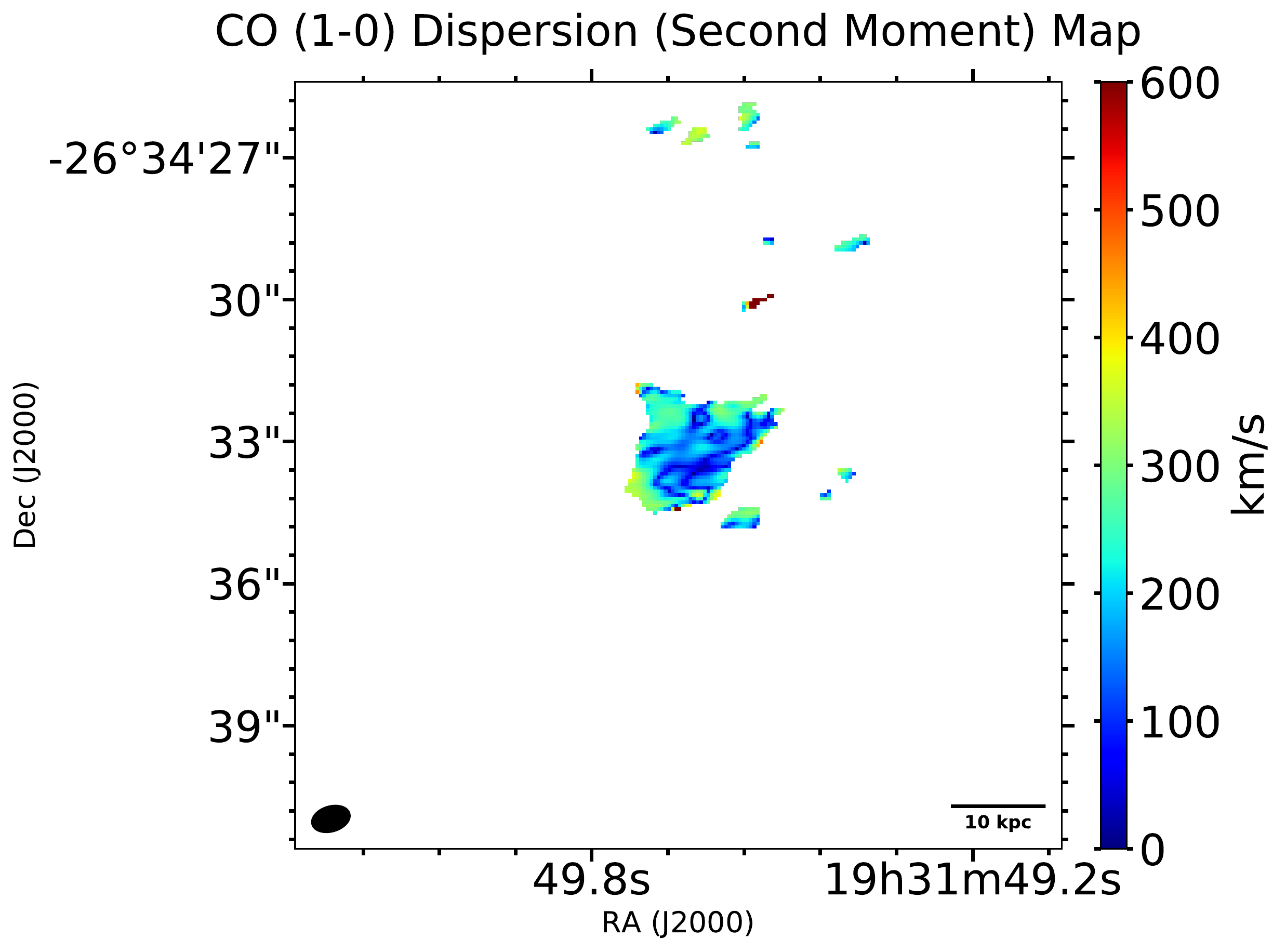}
\includegraphics[height=6.5cm]{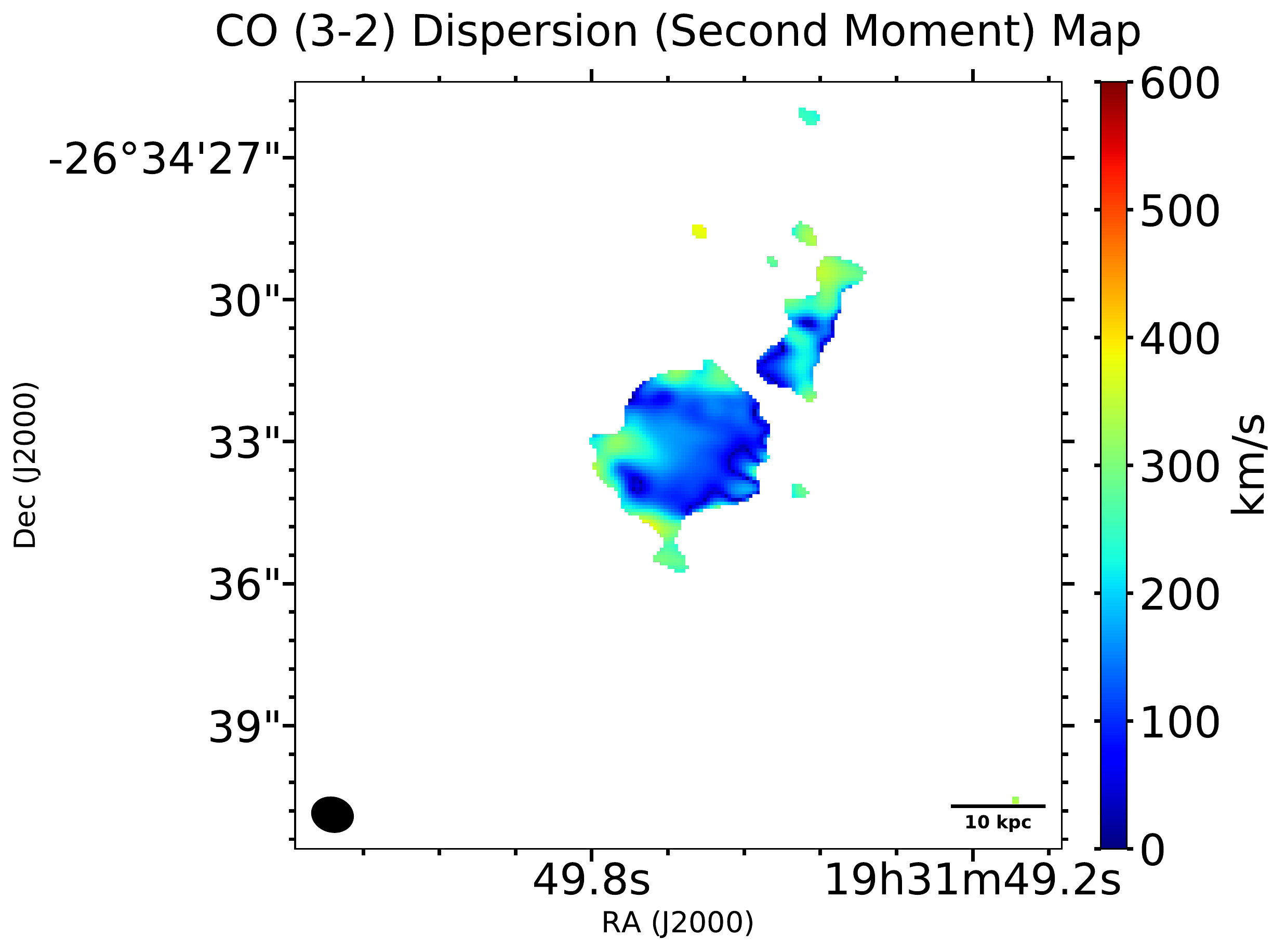}
\end{tabular}
\end{center}
\caption[]
{ \label{fig:Moment_Maps} \textit{Top:} Velocity maps of the CO(1-0) and CO(3-2) molecular gas, computed from the first moment analysis of the datacubes. Images are masked to exclude regions with $< 3\sigma$ flux detections. 
\textit{Bottom:} Velocity dispersion maps computed from second moment analysis of the CO(1-0) and CO(3-2) datacubes. Images are masked to exclude regions with $< 3\sigma$ flux detections. 
}
\end{figure*}

The velocity structure of the molecular gas is complex. We observe a systematic offset of $\sim 300$ km\,s$^{-1}$ between the gas in the tail and the gas in the BCG core. This offset is clearly seen in the velocity map of CO(3-2) (see Figure \ref{fig:Moment_Maps}) and is consistent with the relative velocities between the BCG core and tail obtained by fitting both CO(1-0) and CO(3-2) in the {\it uv}-tapered datacubes. In the MACS 1931 BCG the molecular gas shows no clear morphological symmetry around the core or connection to X-ray cavities that would suggest jet-driven molecular gas outflows like those seen in Abell 1835 and Abell 1664 \citep{2014McNamara_A1835, 2014Russell_A1664}. \cite{2012HL_Cavities} detects probable cavities in the core of MACS1931, however; these detected features are found to the east and west of the BCG, as opposed to the approximately north-south orientation of the extended molecular gas.

\begin{figure*}
\begin{center}
\begin{tabular}{c}
\includegraphics[height=16cm]{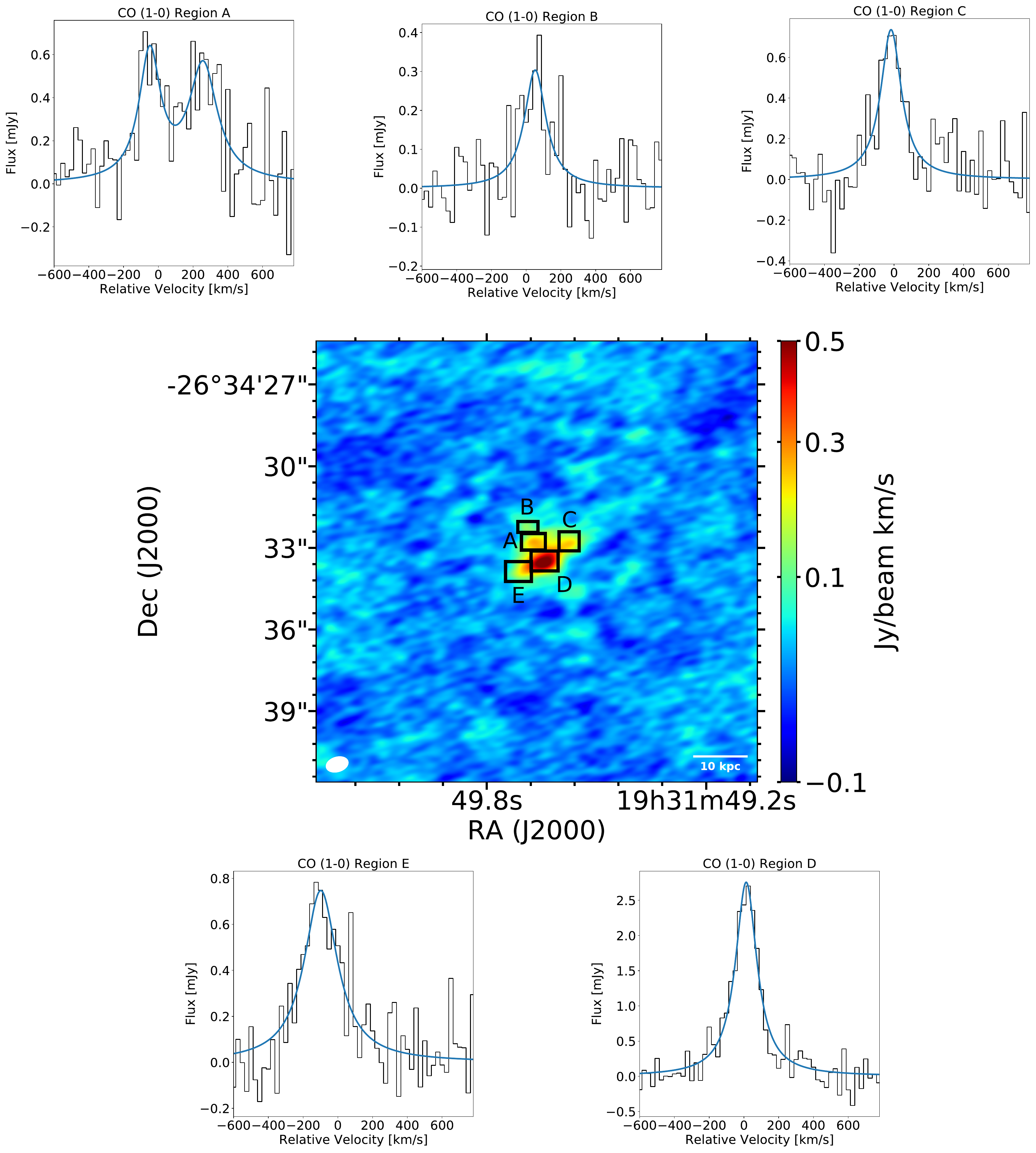}
\end{tabular}
\end{center}
\caption[]
{ \label{fig:Knots_Spectra} CO(1-0) emission spectra for individual knots. The extraction boxes for knots labelled A-E are shown in the intensity image in the center. Boxes were chosen manually to encompass each knot. Best-fit properties of the knots are given in Table \ref{table:Knots_Fits}. All knots are fit with a single-component Lorentzian profile except knot A, which was fit with a two-component model. 
}
\end{figure*}

Molecular gas in the core is concentrated in bright knots (which we have divided up into regions A-E in Figure \ref{fig:Knots_Spectra}) with CO(1-0) velocities ranging from $-98\pm11$ to $258\pm18$ km\,s$^{-1}$ (see Table \ref{table:Knots_Fits}). One of these knots has two velocity peaks, at $-49\pm14$ and $258\pm18$ km\,s$^{-1}$, so it is either rotating or consists of two knots projected on top of each other along the line of sight. Spectra for these knots are shown in Figure \ref{fig:Knots_Spectra}, along with the best-fit Lorentzian profiles. The CO(3-2) and CO(1-0) velocities are consistent on the positions of the knots (see Figure \ref{fig:Moment_Maps}).

\begin{table*}%[]  
\footnotesize  
\caption{CO Knot Properties}
\label{table:Knots_Fits}  
\vspace{1mm}  
\centering  
{  
\begin{tabular}{c|c|r|r}  
             & Integrated      &             &      \\
             &     Flux        & Velocity    & FWHM \\
Region$^{a}$ & (mJy km s$^{-1}$) & (km s$^{-1}$) & (km s$^{-1}$) \\
\hline  
\hline  
A~~ & $139\pm50$         & $-49\pm14$   & $150\pm45$  \\
A$^b$  & ($168\pm59$)  & ($258\pm18$) & ($199\pm58$) \\
B~ & $72\pm24$ & $52\pm14$ & $151\pm39$ \\
C~ & $167\pm40$ & $-17\pm9$ & $145\pm26$ \\
D~ & $651\pm83$ & $13\pm4$ & $150\pm10$ \\
E~ & $273\pm54$ & $-98\pm11$ & $232\pm32$ \\
\end{tabular}  
\begin{flushleft}
$^{a}$ Region boundaries are shown by the rectangular apertures in Figure \ref{fig:Knots_Spectra}.\\
$^{b}$ The spectrum of region A shows two distinct components; the more redshifted component is reported in the parentheses.
\end{flushleft}  
}  
\end{table*}

\begin{figure*}
\begin{center}
\begin{tabular}{c}
\includegraphics[height=6.5cm]{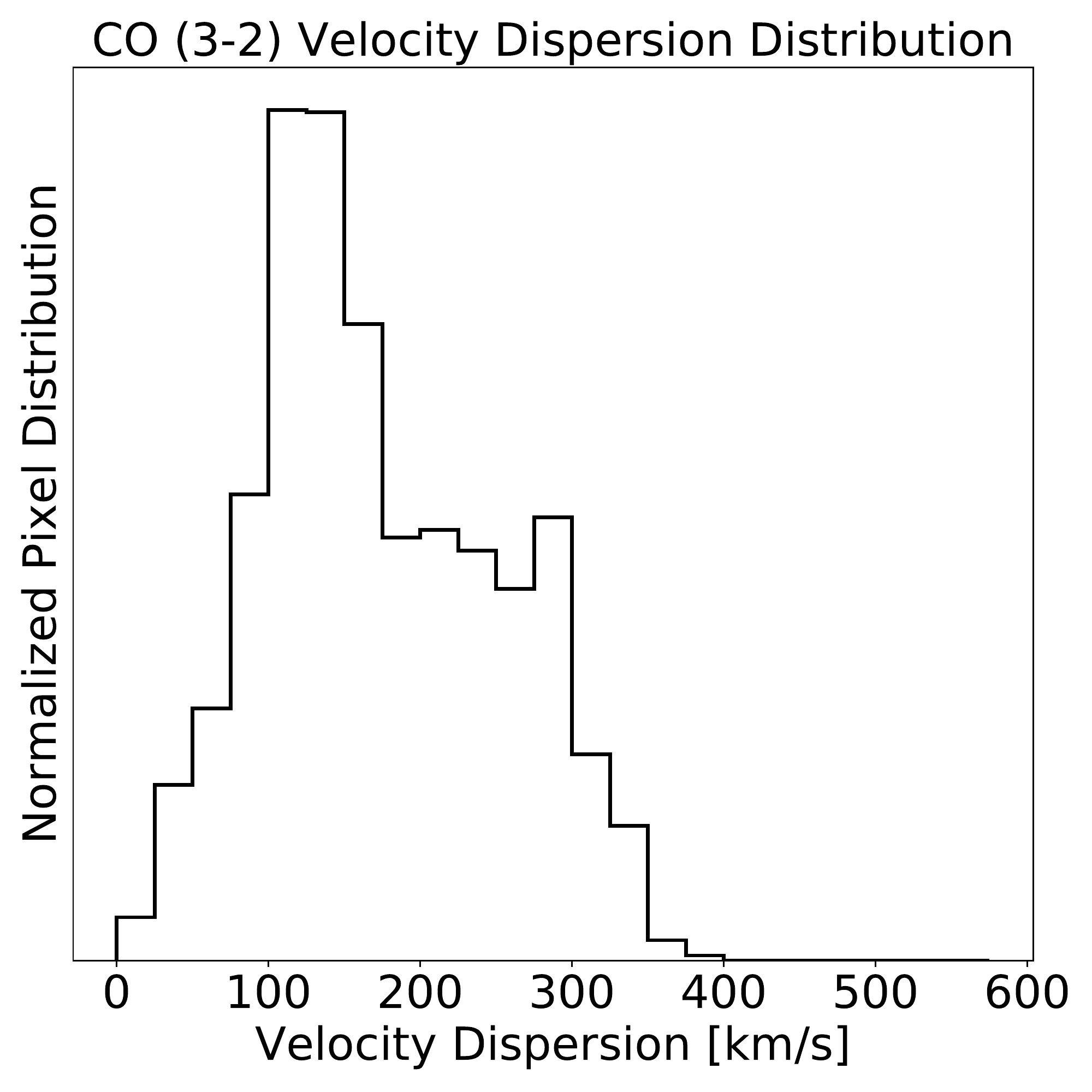}
\includegraphics[height=6.5cm]{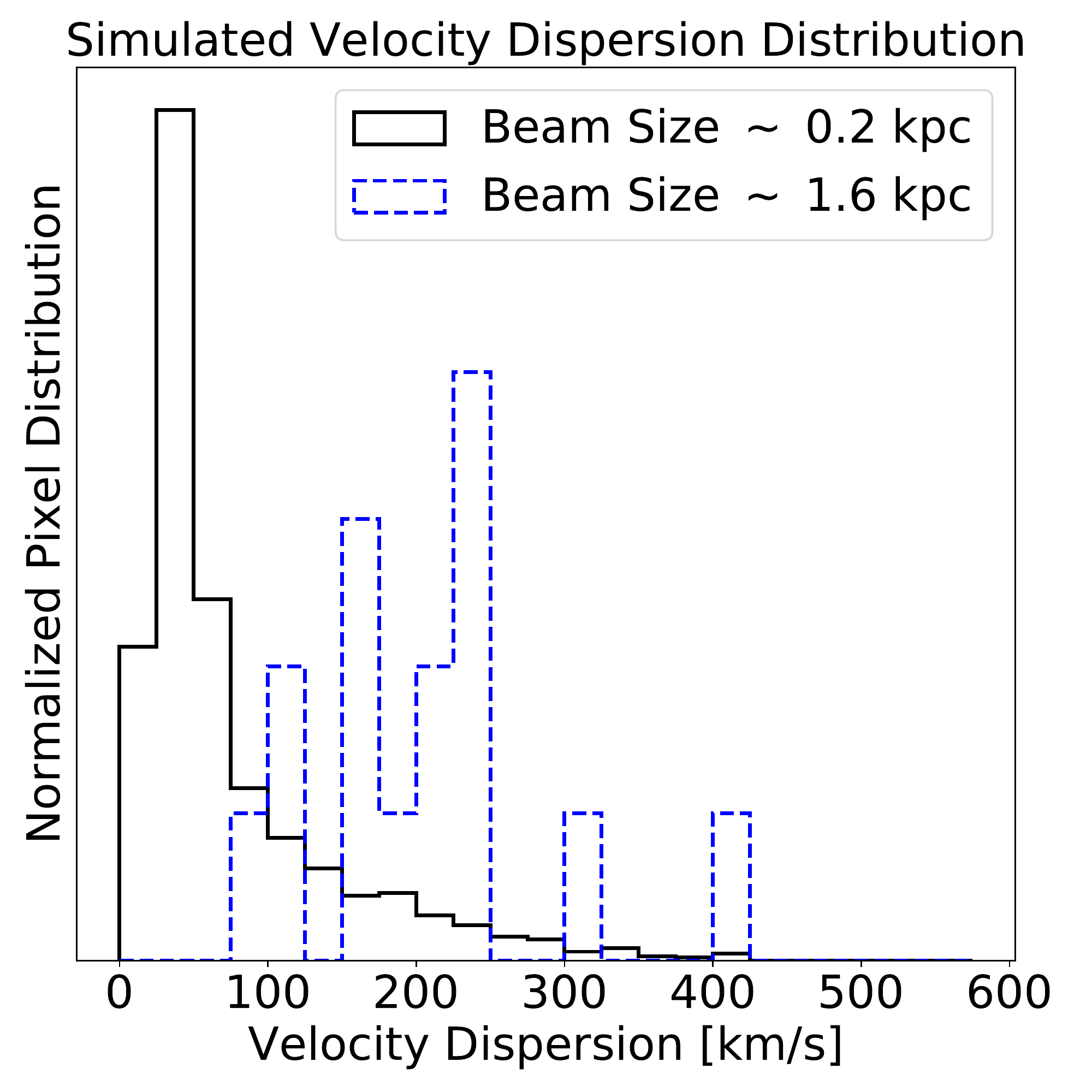}
\end{tabular}
\end{center}
\caption[]
{ \label{fig:VelDisp_Histograms} \textit{Left:} Distribution of CO(3-2) velocity dispersions. \textit{Right:} Distribution of velocity dispersions in a simulated cluster \citep{2015Li_SFAGN}, smoothed with either 0.2 kpc or 1.6 kpc beams. Beam sizes correspond to beams of either $0\farcs04$ or $0\farcs32$ at $z = 0.3525$. 
}
\end{figure*}

The velocity dispersion maps of CO(1-0) and CO(3-2), estimated from analysis of the second moments of the emission lines, are shown in Figure \ref{fig:Moment_Maps}. Velocity dispersion distributions for the two lines are similar, and range between $\sim 25$ km\,s$^{-1}$ and $\sim 400$ km\,s$^{-1}$.  The range of velocity dispersions we observe is similar to observations of CO in other BCGs \citep{2018Vantyghem_RXJ1504, 2016Vantyghem_2A, 2016Russell_PKS0745, 2014Russell_A1664}. Figure \ref{fig:VelDisp_Histograms} shows the distribution of the observed CO(3-2) velocity dispersions alongside predictions for simulated BCGs with conditions similar to MACS 1931 when the simulated data is smoothed to spatial resolutions approaching the resolution of our ALMA data.  The physical beam sizes of our CO images are $4.1 \times 2.6$ kpc for the CO(1-0) image and $4.3 \times 3.6$ kpc for the CO (3-2) image. As the beam size used to smooth the simulated data increases and approaches the beam size of the actual observed data, the typical velocity dispersion measured in the simulated dataset comes into agreement with our observations.

Individual giant molecular clouds in the Milky Way have typical sizes of 5 to 200 pc \citep{2011norman_MW}. Therefore, the angular resolution of our ALMA observations is insufficient to probe the internal velocity dispersion of individual molecular clouds. Instead, the velocity dispersions we measure are likely dominated by random bulk motion of different clouds that are spatially close to one another on scales smaller than our beam size. In the nearby Virgo cluster, the ALMA-measured velocity dispersion of the molecular cloud in M87 is $27 \pm 3$ km\,s$^{-1}$ \citep{2018M87}. This is likely still under-resolved as the velocity dispersion of resolved molecular clouds in the Milky Way and nearby star forming galaxies is typically less than 10 km\,s$^{-1}$ (e.g., \cite{cloudvdisp1984}, \cite{cloudvdisp1998}).

Figure~\ref{fig:Moment_Maps} and Figure~\ref{fig:VelDisp_Histograms} nonetheless show evidence for structure in the velocity dispersion of CO. Dispersions are lowest near the center of the core (i.e. near the position of knot ``D'' in Figure \ref{fig:Knots_Spectra}), and increase towards both the southern and northern peripheries of the molecular gas nebula. Inspection of the 1-D spectra at several locations with high velocity dispersions shows evidence of multiple velocity peaks, suggesting that the increased dispersion is almost certainly due to multiple velocity components along the line of sight, possibly as a result of the molecular clouds having a lower filling factor near the outskirts of the CO emission. 

We estimate the total molecular gas mass of MACS 1931 to be $1.9 \pm 0.3 \times 10^{10}$ M$_{\odot}$ based on the CO(1-0) intensity in Jy km s$^{-1}$, $I_{CO}$. We calculated the gas mass using the relation
\begin{equation}
M_{mol} = 1.05\times 10^{4} \frac{X_{CO}}{2\times 10^{20} \frac{\textrm{cm}^{-2}}{\textrm{K km s}^{-1}}}\frac{D_{L}^{2}}{1+z}I_{CO}
\end{equation}
where $D_{L}$ is the luminosity distance in Mpc and $X_{CO}$ is the CO-to-H$_{2}$ conversion factor in units of $\frac{\textrm{cm}^{-2}}{\textrm{K km s}^{-1}}$ \citep{2013Bolatto_XCO}. We adopted the value $X_{CO} = 0.4\times 10^{20} \frac{\textrm{cm}^{-2}}{\textrm{K km s}^{-1}}$ for starbursts and ULIRGs used in \cite{2016Russell_Phoenix} to estimate the molecular gas mass in the Phoenix cluster, in order to be consistent when comparing the molecular gas masses of BCGs (by adopting the Galactic value of $X_{CO}$, as was done in \cite{2014McNamara_A1835}, our estimate of the molecular gas mass rises to $9.4\pm1.3\times10^{10}$ M$_{\odot}$). The gas reservoir we observe in MACS 1931 is approximately equal to the reservoir in the Phoenix cluster ($2.1\pm0.3 \times 10^{10}$ M$_{\odot}$, \cite{2016Russell_Phoenix}), which along with a handful of other recent observations (Abell 1835: \cite{2014McNamara_A1835}, RXJ0821+0752: \cite{2017Vantyghem_RXJ0821, 2018Vantyghem_RXJ0821}, RXC J1504–0248: \cite{2018Vantyghem_RXJ1504}), make it among the largest molecular gas masses observed in a cool-core BCG.

\subsection{Dust Continuum Emission}

\begin{figure*}
\begin{center}
\begin{tabular}{c}
\includegraphics[height=6.5cm]{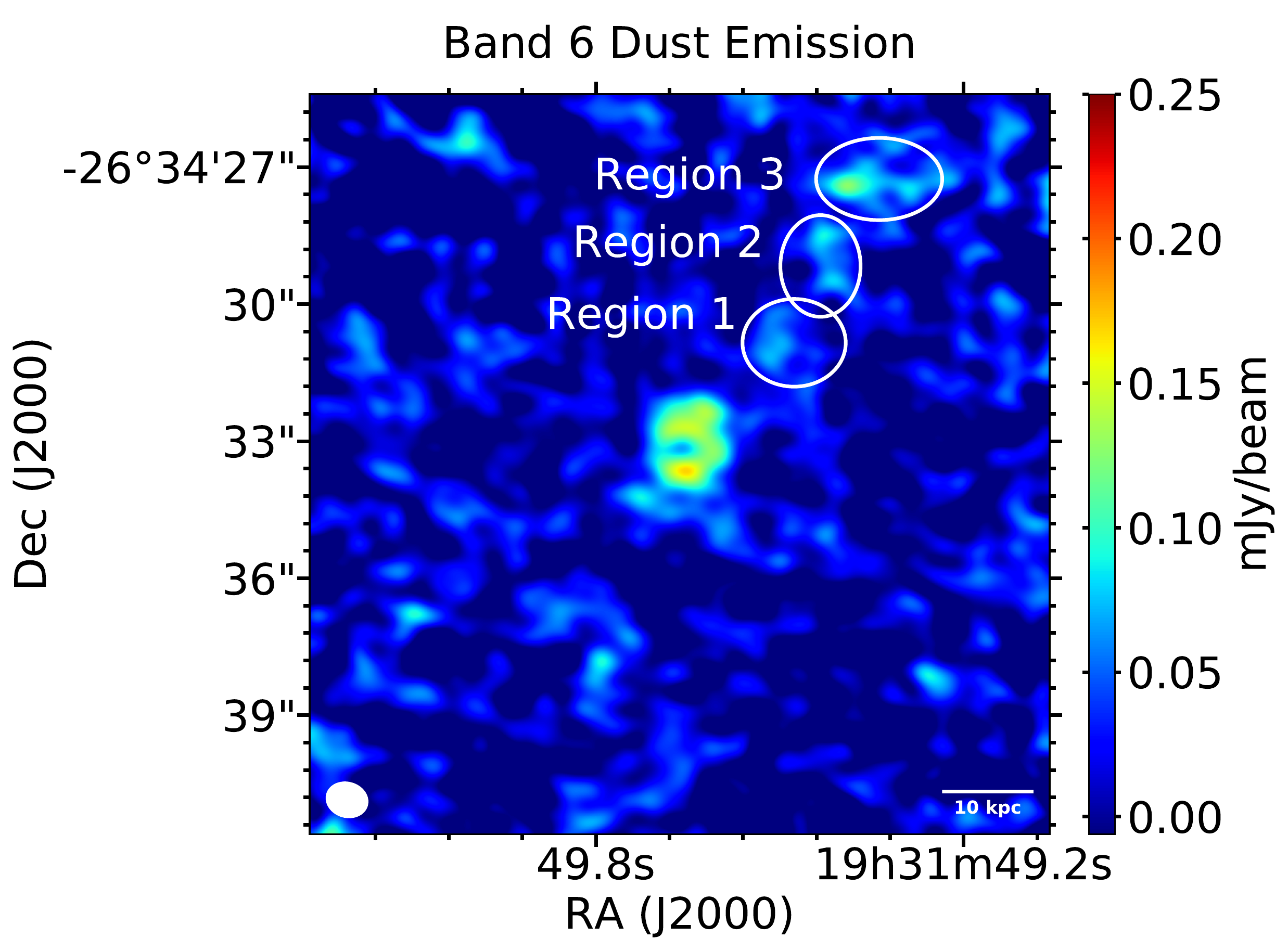}
\includegraphics[height=6.5cm]{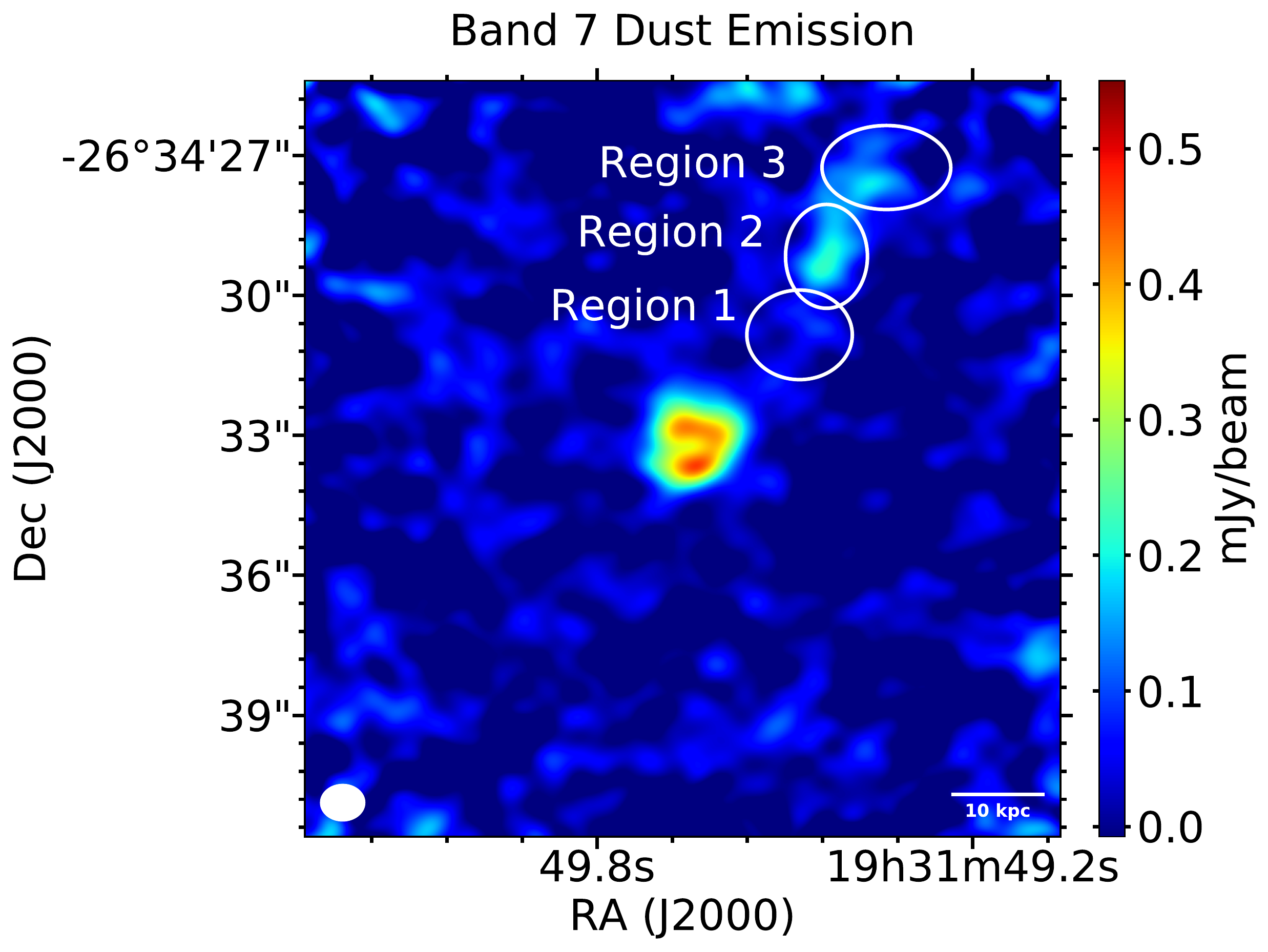}
\end{tabular}
\end{center}
\caption[]
{ \label{fig:Dust_Images} Dust continuum emission in ALMA Bands 6 (336.0 GHz, 892 $\mu$m rest-frame) and 7 (468.5 GHz, 640 $\mu$m rest-frame). The contribution from the AGN point source has been subtracted out in each image. We study the band 7 / band 6 flux ratios in Regions 1-3, represented by the white ellipses, in order to constrain the dust temperature in different parts of the multi-phase gas tail. 
}
\end{figure*}

Images of the dust continuum emission in our Band 6 and 7 observations are shown in Figure \ref{fig:Dust_Images}. These images are shown after performing the AGN point source subtraction described in Section \ref{Sec:Dust_Methods}. Extended dust emission is clearly visible in both bands in the core and tail regions described in Section \ref{sec:CO_Results} above. The Band 6 continuum is centered on 336.0 GHz (892 $\mu$m) in the rest frame and the Band 7 on 468.5 GHz (640 $\mu$m), so the emission we observe in these bands is either due to very cold dust or is the Rayleigh-Jeans tail of the greybody emission of hotter dust. The dust emission in MACS 1931 is most prominent in the BCG core and in the northern tip of the tail, corresponding to the edge of the extended H$\alpha$ and UV emission seen in the CLASH UV-optical images shown in Figure \ref{fig:CO_Images}.

 We checked to ensure that the emission features we observe are not due to free-free emission or a faint, extended synchrotron source. We examined the Band 3 continuum image after convolving it with the Band 6 synthesized beam, since both free-free and synchrotron emission will appear brighter in Band 3. The Band 3 fluxes have detection significances of $\lesssim 0.5\sigma$ in the apertures labelled Regions 1, 2, and 3 in Figure \ref{fig:Dust_Images}. The $3\sigma$ limit in Band 3 is 0.023 mJy/beam, which is about a factor of 3 fainter than the limits in either Bands 6 or 7, implying that any contamination by extended free-free or synchrotron emission in our dust images is below the RMS noise in Bands 6 and 7.

We also estimated the total free-free and synchrotron emission in Band 6 due to star formation using the relationship derived in \cite{2006Schmitt_RadioSFR}. A SFR of 250 M$_{\odot}$ yr$^{-1}$ yields a total flux in Band 6 of 0.08 mJy. If this flux is evenly distributed throughout the region where Band 6 emission is detected at $\geq 3 \sigma$, we predict a Band 6 flux due to star formation of 0.01 mJy/beam, which is well below the RMS noise of 0.020 mJy/beam. The predicted Band 3 flux, meanwhile, is 0.014 mJy/beam, which is consistent with being below the detected limit in our Band 3 image. Since the dust emission is higher in Band 7 than Band 6, while the free-free and synchrotron component is lower, we did not perform the same check for Band 7.

Since we have multiple bands of ALMA FIR photometry, we are able to constrain the dust temperature using a simple model for the flux. Given the long rest-wavelengths of our Bands 6 and 7 photometry, we can only reliably derive the temperature of very cold $T_{d} \lesssim 30$ K dust from the ALMA data alone, since emission from hotter dust in these bands is proportional to both the dust luminosity and $T_{d}$. However, we can still place constraints on the temperature of the cold dust in the tail, especially in regions with relatively high flux in Band 6.

\cite{2012Casey_Dust} describes a mid-IR powerlaw + greybody model that represents the non-PAH continuum emission due to dust in a ULIRG-like environment with the expression
\begin{equation}\label{eq:Dust}
    S\left(\lambda\right) = N\frac{\left(c/\lambda\right)^{3}\left(1-e^{-\left(\lambda_0/\lambda\right)^{\beta}}\right)}{e^{hc/\lambda kT_d}-1} + N_{pl}\lambda^{\alpha}e^{-\left(\lambda/\lambda_{c}\right)^{2}},
\end{equation}
where $S\left(\lambda\right)$ is the flux density in Jy, $\lambda_{c}$ is the turnover wavelength separating the regimes where the mid-IR powerlaw and the greybody dominate the emission, $\lambda_{0} \equiv 200$ $\mu$m, $N$ is the overall normalization, $T_d$ is the dust temperature, $\alpha$ is the mid-IR powerlaw index, and $\beta$ is the greybody emissivity index. Both $\lambda_{c}$ and the powerlaw normalization, $N_{pl}$, are a function of the other parameters in the model \citep{2012Casey_Dust}. Given the long wavelengths of our two ALMA bands, we choose to neglect the mid-IR powerlaw component of the model in Equation \ref{eq:Dust} and model the dust emission as a simple greybody.

Since the dust components we will be able to constrain with this method are cold, we need to account for the impact of the cosmic microwave background (CMB) on our observations, which can potentially bias our estimated temperature constraints by several K. We note, however, that this correction does not qualitatively change the results of our analysis. We modify the first term in Equation \ref{eq:Dust} to be:
\begin{equation}\label{eq:ModDust}
\begin{aligned}
    S\left(\lambda\right) = N\left[\frac{\left(c/\lambda\right)^{3}\left(1-e^{-\left(\lambda_0/\lambda\right)^{\beta}}\right)}{e^{hc/\lambda kT_d}-1}  \right. \\ \left. + \frac{\left(c/\lambda\right)^{3}e^{-\left(\lambda_0/\lambda\right)^{\beta}}}{e^{hc/\lambda kT_{CMB}}-1}\right],
\end{aligned}
\end{equation}
where $T_{CMB} = 2.73\left(1+z\right) = 3.69$ K is the temperature of the CMB at $z = 0.3525$. This equation requires one further modification since the interferometer configuration we use does not have baselines short enough to detect the spatial scales that contain the CMB power. Therefore, while we observe the dust absorption of the CMB, we do not observe the CMB itself, and need to subtract an additional term of $\left(c/\lambda\right)^{3} \left( e^{hc/\lambda kT_{CMB}}-1\right)^{-1}$ from equation \ref{eq:ModDust} (see e.g. \cite{2016Zhang_HighZDust}). The final expression for the Band 7 to Band 6 continuum flux ratio is therefore
\begin{equation}\label{eq:Dust_Corrected}
\begin{aligned}
    \frac{S\left(\lambda_{7}\right)}{S\left(\lambda_{6}\right)} = {} & \left(\frac{\lambda_{6}}{\lambda_{7}}\right)^{3}\left(\frac{1-e^{-\left(\lambda_0/\lambda_7\right)^{\beta}}}{1-e^{-\left(\lambda_0/\lambda_6\right)^{\beta}}}\right)\\
    & \times \left(\frac{\frac{1}{e^{hc/\lambda_{7}kT_d}-1} - \frac{1}{e^{hc/\lambda_{7}kT_{CMB}}-1}}{\frac{1}{e^{hc/\lambda_{6}kT_d}-1} - \frac{1}{e^{hc/\lambda_{6}kT_{CMB}}-1}}\right),\\
\end{aligned}
\end{equation}
where $\lambda_{7}$ is the rest wavelength of the Band 7 continuum ($640\, \mu$m) and $\lambda_{6}$ the rest wavelength of the Band 6 continuum ($892\, \mu$m).

\begin{figure*}
\begin{center}
\begin{tabular}{c}
\includegraphics[height=8.5cm]{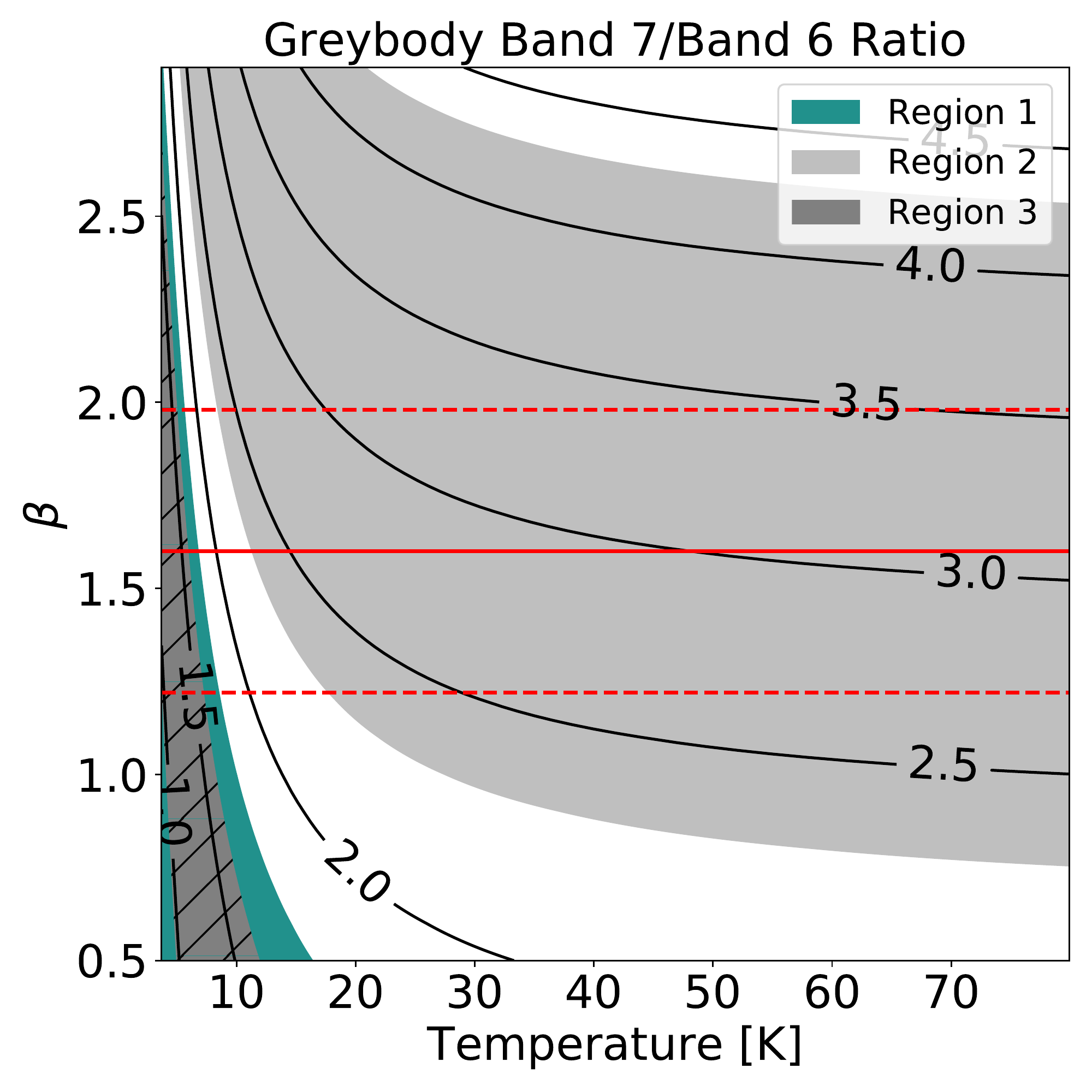}
\includegraphics[height=8.5cm]{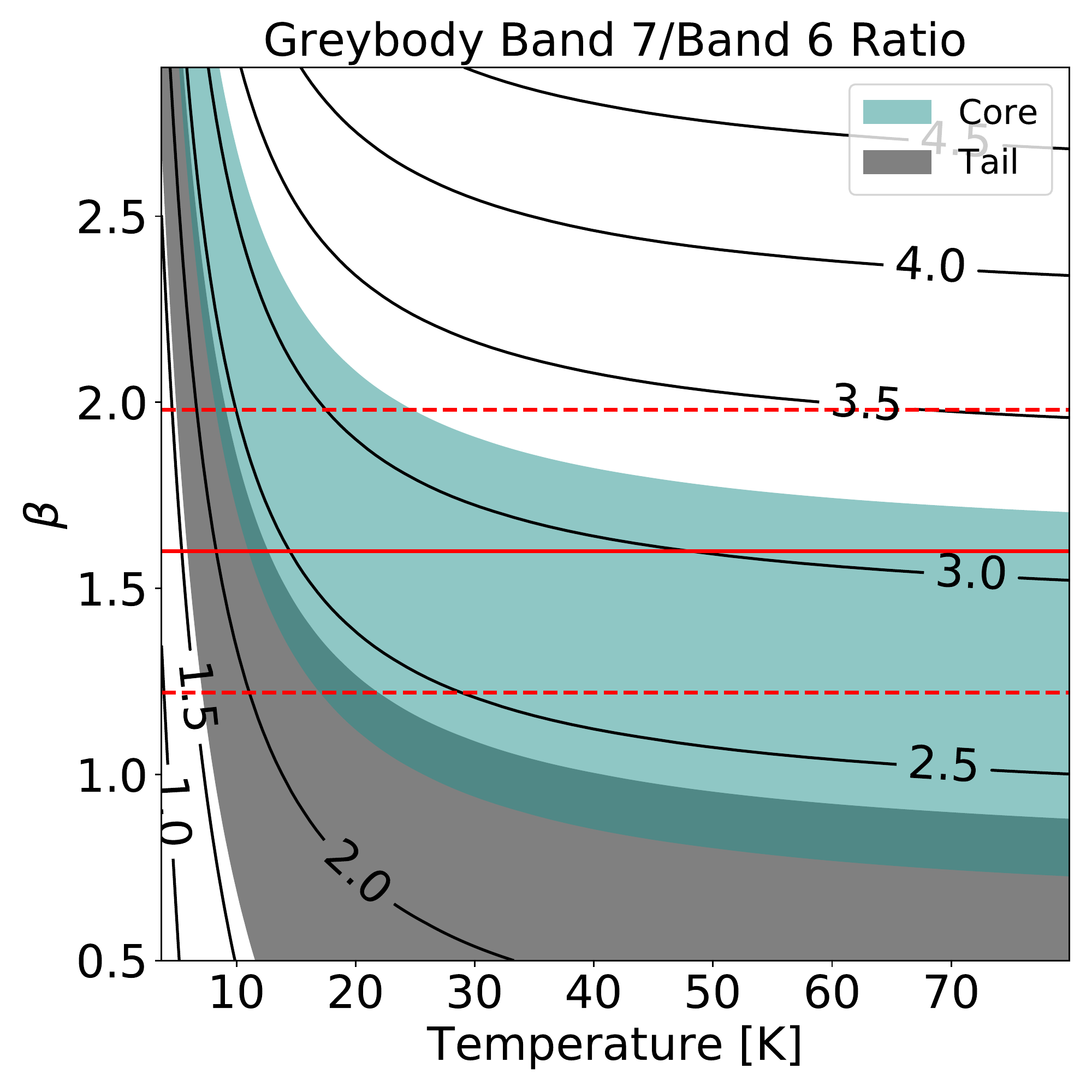}
\end{tabular}
\end{center}
\caption[]
{ \label{fig:TBeta_Analysis} Predicted Band 7 to Band 6 flux ratios for a greybody emission spectrum are shown as a function of dust temperature and the greybody emissivity index $\beta$. The shaded-in regions in the left-hand plot bound the $1\sigma$ uncertainties for the flux ratios measured in Regions 1-3 shown in Figure \ref{fig:Dust_Images}. In the right-hand plot, the shaded in regions show the ratios for the core and tail, where we measured fluxes for the tail in the box shown in Figure \ref{fig:Tail_SB}. The solid red line denotes $\beta = 1.6$, the value for ULIRGs measured in \cite{2012Casey_Dust}, and the red dashed lines denote the $\pm0.38$ uncertainty in that value. 
}
\end{figure*}

The contour plot in Figure \ref{fig:TBeta_Analysis} shows the predicted values of the flux ratio as a function of $T_d$ and $\beta$. The parameter $\beta$ depends on the impact of the physical properties of the dust on its emissivity and for local sources has a value between 0.8 and 2.4 \citep{2003Dupac_Pronaos}. \cite{2012Casey_Dust} measured $\beta = 1.6\pm0.38$ for a sample of ULIRGs, which we take to be the most plausible range for $\beta$ in the MACS 1931 BCG.

We examined three regions outside the core that are relatively bright in Band 6, denoted by the solid ellipses in Figure \ref{fig:Dust_Images} and labelled Regions 1-3. We smoothed the Band 6 and 7 images to a common beam, and measured the flux ratios in these regions. The Band 7 to Band 6 flux ratios for these regions are $1.25\pm0.53$, $3.29\pm0.99$, and $1.29\pm0.33$, respectively. We also measure the flux ratio in the BCG core, obtaining a value of $2.74\pm0.46$, and the overall ratio in the tail, finding a value of $2.00\pm0.40$.

Shaded bands showing the $1\sigma$ intervals around the ratios for Regions 1-3 are shown in Figure \ref{fig:TBeta_Analysis}. There is a very cold dust component in Regions 1 and 3 --- for $1.22 \leq \beta \leq 1.98$, we estimate $T_d \lesssim 8.6$ K, and $\lesssim 7.3$ K at the $1\sigma$ confidence level, and $\lesssim 17.3$ K, and $\lesssim 10.2$ K at the $95\%$ confidence level respectively. If $\beta = 1$, the temperatures for these regions are $\lesssim 26.4$ K and $\lesssim 12.5$ K at $95\%$ confidence.

A lower limit of 8.4 K can be placed on the dust temperature of Region 2. If we assume $\beta = 1.6$, then we can constrain $T_d$ to $> 11.2$ K.  The measured value of the flux ratio of $3.29$ in Region 2 implies a temperature of 30.5 K if $\beta = 1.98$, but $>70$ K once $\beta < 1.8$, so depending on the emissivity the dust in this region may either be relatively cold or warm.

In the core of the BCG, if we assume $\beta = 1.6$, we can place a constraint of $T_d > 11.0$ K given the $1\sigma$ errorbars on the flux ratio. However, the tail seems to be substantially colder overall; we find that $5.0 < T_d < 21.9$ K for $\beta = 1.6\pm0.38$. 

We note that when we model and subtract the bright point sources in the Band 6 and 7 continuum images, the peak positions of the best-fit point source in the two bands differ by less than 0.25 pixels, so the low flux ratios we obtain are not likely to be the result of astrometric offsets between bands.

\section{Discussion}\label{sec:Discussion}

The MACS 1931 BCG is surrounded by a massive molecular gas reservoir containing $1.9 \pm 0.3 \times 10^{10}$ M$_{\odot}$ of gas. If this gas all condensed in the last $\sim 100$ Myr, putting it on the same timeframe as the duration of the ongoing starburst in this system \citep{2017Fogarty_CLASH}, then this would imply that gas has to be cooling out of the ICM at a mean rate of $\sim 190$ M$_{\odot}$ yr$^{-1}$. \cite{2011Ehlert_MACS1931} attempted to estimate the cooling rate of the ICM in the core of MACS 1931 using \textit{Chandra} ACIS observations and found a rate of $165^{+45}_{-67}$ M$_{\odot}$ yr$^{-1}$. Assuming the value measured by \cite{2011Ehlert_MACS1931} is a reasonable upper limit for the actual rate of cooling out of the X-ray phase, this measurement implies either that molecular gas in the core of MACS 1931 has been forming at about the rate predicted by X-ray observations, or MACS 1931 began forming gas before it began forming stars.

Previously, we showed that the BCG contains an extreme and relatively young starburst-- in \cite{2017Fogarty_CLASH} we found\footnote{In \cite{2017Fogarty_CLASH}, we report the MACS 1931 BCG SFR and burst duration values in log units, reflecting the approximately log-normal distributions of the best-fit probability density functions. The best-fit values quoted here correspond to the modes of these parameters assuming a log-normal distribution for each.} that for the MACS 1931 BCG, the SFR = $240^{+197}_{-70}$ M$_{\odot}$ yr$^{-1}$ and we constrained the duration of the starburst to be $74^{+145}_{-30}$ Myr. The SFR of the MACS 1931 BCG coupled with our estimate of the molecular gas mass imply the molecular gas has a depletion time of $\sim 80$ Myr, suggesting it is about a factor of 10 shorter than the typical $\sim$ Gyr depletion times inferred in \cite{2008ODea_BCGSF} for cool-core clusters, but about a factor of $\sim 2-3$ longer than the $\lesssim 30$ Myr timescale for depleting the gas in the Phoenix cluster \citep{2014McDonald_PhoenixGas}. In star forming galaxies (that are usually not BCGs), observations suggest that gas depletion time is shorter (star formation efficiency is higher) when SFR is higher, especially in starburst galaxies \citep{Kennicutt2012}. \citet{2011Voit_BCGMassLoss} show that BCGs follow a similar trend. Our observations of a starburst BCG reveals a higher star formation efficiency than a normal star-forming BCG. This suggests that the physical processes governing star formation are likely similar in galaxies of all sizes.

In \cite{2017Fogarty_CLASH} we discussed the possibility that MACS 1931 is transitioning from a rapidly star forming stage, similar to the Phoenix cluster, to a lower but steadier star forming stage, similar to the majority of cool-core clusters with BCG SFRs around $1-10$ M$_{\odot}$ yr$^{-1}$ and Gyr-scale gas depletion times. In addition to having an intermediate depletion time, there are several lines of reasoning to support this hypothesis. MACS 1931 is like the Phoenix cluster as it possesses a large molecular gas reservoir that probably formed under the influence of an atypical X-ray bright BCG AGN. However, in the case of the Phoenix cluster, molecular gas is observed in prominent ridges outlining the boundaries of radio bubbles, suggesting ongoing formation at these sites \citep{2016Russell_Phoenix}. This morphological correspondence is also seen in clusters such as Abell 1795 with more modest SFRs, suggesting this formation process occurs on various orders of magnitude \cite{2017Russell_A1795}. 

While \textit{Chandra} observations reveal the presence of structures in the ICM consistent with cavities, these features are not as distinct as they appear in lower redshift BCGs, where radio plasma from collimated jets expand to fill X-ray cavities with clearly defined boundaries \citep{2012HL_Cavities, 2007McNamara_AGNFeedback}. Clear evidence of a cavity 20.6 kpc to the West is present, with weaker evidence for a corresponding cavity 29.5 kpc to the East. This orientation is consistent with observations of the 1.5 GHz continuum, which is elongated along an east-west axis \citep{2011Ehlert_MACS1931, 2018Yu_JVLA}. Although clear interfaces between the cavities and the surrounding ICM are not observed, the orientation of the AGN jet implied by the cavities is orthogonal to the orientation of the multiphase gas reservoir we observe. Unlike the molecular gas in the Phoenix cluster and others, the morphology and velocity structure of the molecular gas in MACS 1931 is not consistent with having recently formed from material along the jet axis or along cavity interfaces. The tail is within a few kpc of the possible northern edge of the western cavity, but does not correspond with the possible boundary of the AGN-excavated cavity (see Figure \ref{fig:Discussion_Figure}), although very faint ($\lesssim 2\sigma$) emission in the {\it uv}-tapered CO(3-2) image (see Figure \ref{fig:CO_Images}) suggests a small amount of gas may be present closer to the cavity. Our results suggest that the molecular gas structure we observe may have originally formed around the western cavity implied by \cite{2012HL_Cavities}, but much of it must have since migrated to its present position. Furthermore, if the features observed in the X-ray are not cavities but are instead a chance morphological alignment of ICM over- and under- densities, this reinforces our supposition that the molecular gas in MACS 1931 must not have formed recently at X-ray cavity interfaces. In this case, initial gas formation will need to have occurred long enough ago for the jet-inflated X-ray cavities to have dissipated.

It is difficult to say definitively whether the molecular gas in the tail is inflowing or outflowing from the BCG core, although it is clearly exhibiting significant motion along the line of sight. However, given the morphological clues presented by our observations, we suspect that the tail is an infalling feature. As noted above, the tail does not lie along the axis connecting the jet-excavated cavities reported in \cite{2012HL_Cavities}, and does not have the bi-modal symmetry characteristic of jets. Given that a power source such as a wind would need to propel the material along a preferred direction through a relatively narrow solid angle in order to produce the observed tail feature, it is more plausible that the material is infalling. In simulations, complex arcs of gas fall back into the centers of BCGs after initially being propelled outwards by AGN jets and cavities. Our observations of MACS 1931 exhibit features that are similar to those seen in simulations \citep{2014Li_ColdClumps, 2015Li_SFAGN}.

The balance of evidence for MACS 1931 leads us to believe that the AGN in the BCG of this cluster recently experienced an energetic outburst similar to that observed in the Phoenix cluster, possibly as a result of temporarily enhanced AGN accretion. If this outburst temporarily stimulated rapid condensation and star formation due to the uplift of low-entropy ICM plasma over the past $\sim 100$ Myr, the result may be what we observe\,-- a BCG starburst of several 100 M$_{\odot}$ yr$^{-1}$ being fueled by a reservoir of molecular gas that was originally formed, or began forming, during prior uplift and is now cycling back down towards the BCG core. 

\subsection{CO Excitation Mechanisms}

We are able to place constraints on the properties of the molecular gas by studying the ratios of CO(1-0), (3-2), and (4-3). The molecular gas line ratios are defined as 
\begin{equation}
R_{ij} = \frac{I_{\mbox{\scriptsize CO (i - (i-1))}}}{I_{\mbox{\scriptsize CO (j - (j-1))}}}\frac{j^{2}}{i^{2}}.
\end{equation}
Line ratios $R_{ij}$ are equivalent to line brightness temperature ratios, and can be used to determine the level of excitation of CO \citep{2009Dannerbauer_CO, 2005Solomon_CO}. $R_{ij} = 1$ implies the relative excitation of two lines is consistent with thermal excitation \citep[e.g.][]{2015Daddi_CO}. Based on the fluxes in Table \ref{table:Line_Params}, we find that for MACS 1931 as a whole, $R_{31} = 0.93\pm0.16$ and $R_{41} = 0.61\pm0.10$. $R_{31}$ in MACS 1931 is similar to the $R_{31} \sim 0.8$ values commonly observed in low-redshift BCGs (e.g., \cite{2001Edge_MolecularGas, 2016Russell_PKS0745, 2017Vantyghem_RXJ0821}).  

\begin{figure*}
\begin{center}
\begin{tabular}{c}
\includegraphics[height=12cm]{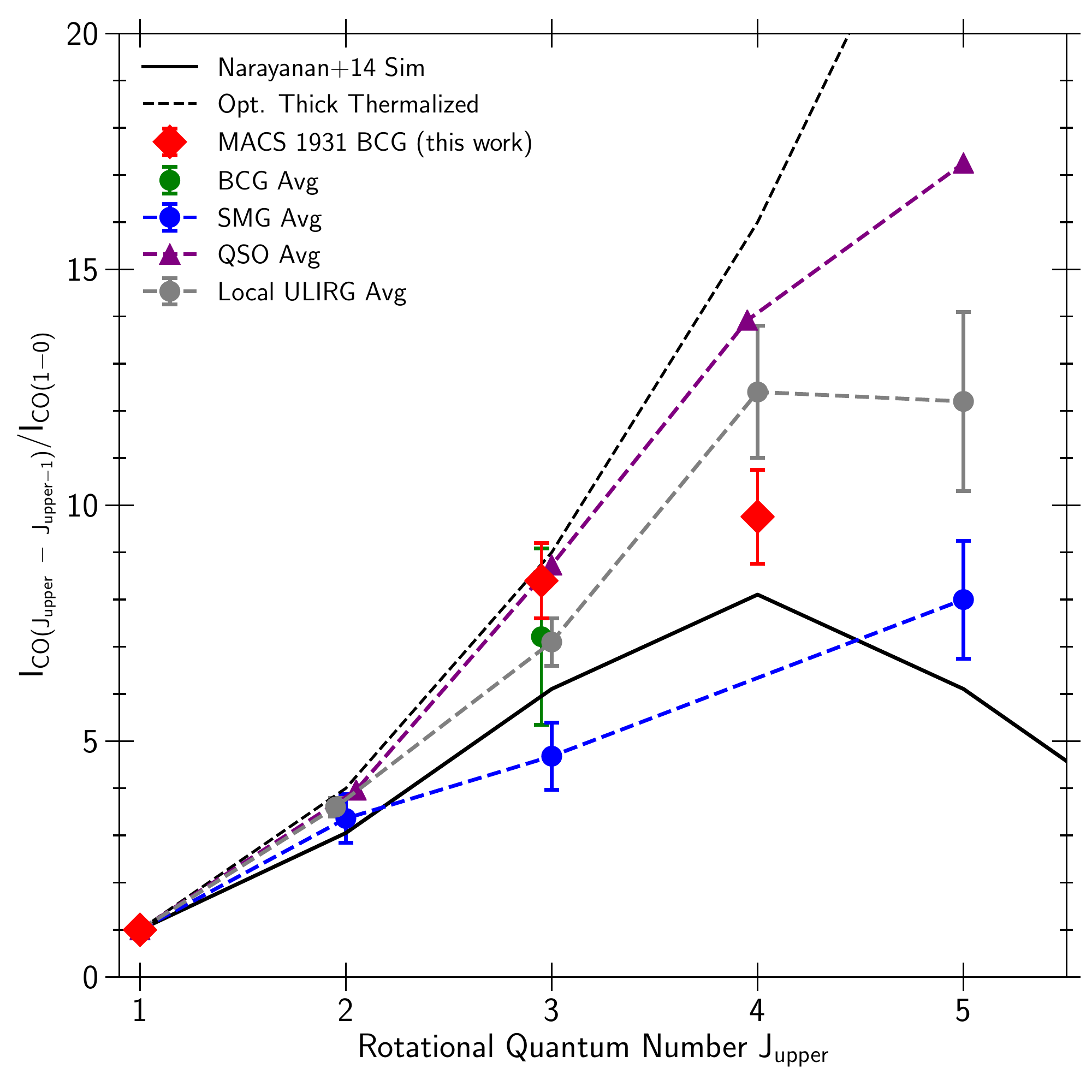}
\end{tabular}
\end{center}
\caption[]
{ \label{fig:CO_SLED} The CO SLED for the MACS 1931 BCG (red diamonds) as well as those for various other galaxy populations and simulated star-forming galaxies. In this figure, all CO(J$_{\mathrm{upper}}$-J$_{\mathrm{upper}-1}$) intensities for both measured and simulated CO transitions are normalized to CO(1-0) line flux. The dashed black line is the expected CO SLED for an optically thick, thermalized gas. The solid black line is the CO SLED for a starforming galaxy predicted by \cite{2014Narayanan_CO}. See text for further details on the origin of the other data points.
}
\end{figure*}

The relative excitation levels of CO(3-2) and CO(4-3) in MACS 1931 places the excitation level of the molecular gas in this system similar to gas typically seen in low-redshift ULIRGs based on $R_{31}$, while $R_{41}$ is below the average for the local ULIRGs \citep{2015Daddi_CO}. The CO spectral line energy distributions (SLEDs) for MACS 1931 and other astrophysical sources are shown in Figure \ref{fig:CO_SLED}. The MACS 1931 values are the red diamonds. The solid black line is the CO SLED for a starforming galaxy predicted by \cite{2014Narayanan_CO}, with CO excited by star formation with a surface density of $\Sigma$SFR = 1 M$_{\odot}$ yr$^{-1}$ kpc$^{-2}$, adapted from \cite{2015Daddi_CO}. For reference, the UV-derived $\Sigma$SFR of the MACS 1931 BCG is $0.83\pm 0.06$ M$_{\odot}$ yr$^{-1}$ kpc$^{-2}$ \citep{2015Fogarty_CLASH}. The SMG (blue points) and local ULIRG averages (grey points) are derived from the samples studied in \cite{2015Daddi_CO} (see also \cite{2012Papadopoulos_ULIRG} and \cite{2013Bothwell_SMG}), while the QSO average (purple triangles) is taken from \cite{2013Carilli_CoolGas}. For our estimate of the BCG CO(3-2) average flux (green point) we used sources in the literature with reported ALMA CO(1-0) and CO(3-2) observations, and adopted the scatter in these observations for the uncertainty. These include RXJ 1504-0248 \citep{2018Vantyghem_RXJ1504}, 2A 0335+096 \citep{2016Vantyghem_2A}, PKS0745-191 \citep{2016Russell_PKS0745}, and Abell 1664 \citep{2014Russell_A1664}.

We compare MACS 1931 to SMGs and ULIRGs since both SMGs and ULIRGs have several characteristics in common with the MACS 1931 BCG. As with SMGs, the BCG is at the high end of the galactic stellar mass distribution and yet the observed large star formation rate contributes, at most, a few percent to their overall stellar mass budgets \citep{2010Michalowski_SMG, 2011Gonzalez_SMG}. SMGs are hypothesized to be the progenitors of elliptical galaxies and are often observed undergoing massive bursts of star formation as large as several 1000 M$_{\odot}$ yr$^{-1}$ \citep{2006Pope_SMG}. Meanwhile, the infrared luminosity of MACS 1931 is $1.4\pm0.2 \times 10^{12}$ L$_{\odot}$, consistent with the definition of a ULIRG \citep{2015Santos_AGN}. While the majority of ULIRGs are fueled by galaxy interactions, the MACS 1931 BCG is typical of a ULIRG in terms of hosting a powerful dusty starburst and AGN \citep{2006Lonsdale_ULIRG, 2015Santos_AGN, 2017Fogarty_CLASH}.

Figure \ref{fig:CO_SLED} shows that there is mild tension between both the observed CO(3-2) and CO(4-3) transitions and the average behavior of the ULIRG sample. However, the MACS 1931 CO SLED is not consistent with the SMG sample. The SLED is also inconsistent with the \cite{2014Narayanan_CO} prediction for molecular gas excited by star formation with a similar $\Sigma$SFR to MACS 1931, although the CO(4-3) transition for MACS 1931 lies between the ULIRG SLED and the \cite{2014Narayanan_CO} prediction.

The tension between the MACS 1931 BCG and the ULIRG sample by itself does not imply a significant discrepancy between the BCG and the ULIRG population, since the CO SLED we plot is an average with considerable variation. However, it is noteworthy that while $R_{31}$ implies that CO(3-2) is consistent with 1, implying extreme excitation conditions, $R_{41}$ is substantially less than this value. \cite{2012Papadopoulos_ULIRG} and \cite{2015Rosenberg_ULIRG} note that the conditions required for the levels of molecular gas excitation that produce $R_{j1} \gtrsim 1$ at the scales of individual molecular clouds exist near extremely UV-bright young stars and supernova shocks, but producing these levels of excitation for an entire ULIRG requires either a very strong X-ray dominated region from a QSO or large scale mechanical turbulence or cosmic ray heating. 

While AGN emission plays a role in the MACS 1931 CO SLED and provides a compelling explanation for the observed CO(3-2) line strength, QSO heating does not explain the average CO(4-3) level we observe. On the other hand, recent mechanical-mode AGN feedback can potentially provide both cosmic ray and turbulent heating to excite the MACS 1931 CO SLED. Comparison of the CO SLEDs suggest that the molecular gas excitation budget in MACS 1931 may be driven by a combination of star formation and processes that, where more strongly present, can drive ULIRG CO excitation ratios to $> 1$ \citep{2012Papadopoulos_ULIRG}. 

A related possibility is that some of the gas in MACS 1931 may be undergoing processes similar to what is observed in NGC 1275. Young stars are found to be associated with H$\alpha$ and molecular filaments in NGC 1275, the central galaxy of Perseus, but there is often an offset of 0.5-1 kpc between young stars and filaments \citep{Canning2014,Li2018}. There are also many filaments in Perseus that are not actively forming stars \citep{2008Salome_NGC1275, 2017Lim_Perseus}.  These filaments may experience pressure support from turbulence and magnetic fields, and heating via infiltration and radiation from the surrounding ICM that suppresses the collapse of this gas into star forming cores \citep{2001Conselice_Perseus, 2008Fabian_PerseusFilaments, 2008Salome_NGC1275, 2011Fabian_PerseusFilaments}. This process may help to drive CO(3-2) excitation but around the denser gas traced by CO(4-3) would potentially need to be weaker, and therefore provide an explanation for the CO SLED in MACS 1931.

\begin{figure*}
\begin{center}
\begin{tabular}{c}
\includegraphics[width=14cm]{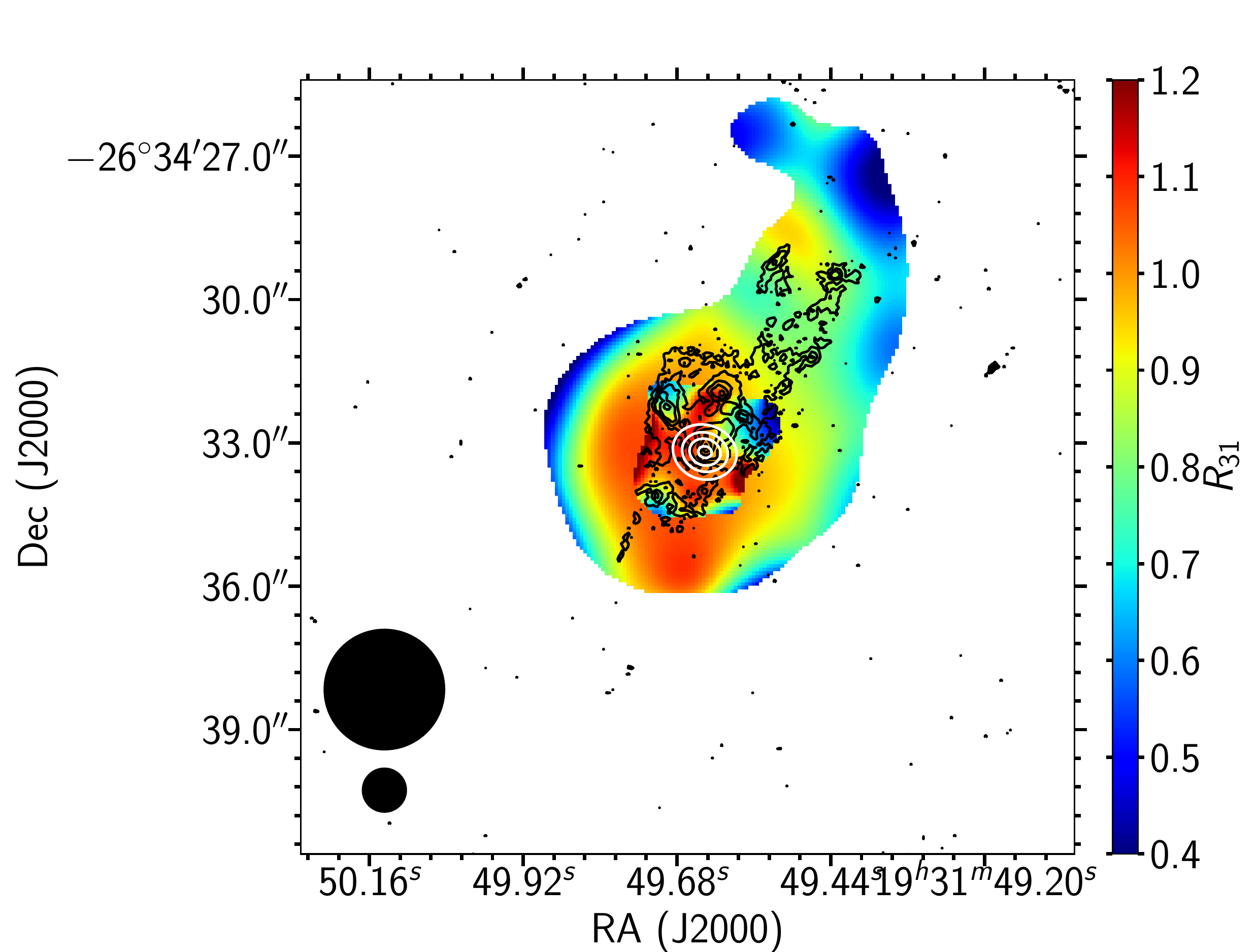}
\end{tabular}
\end{center}
\caption[]
{ \label{fig:R31_Map} Map of $R_{31}$ on the plane of the sky for the MACS 1931 BCG. The map shows all pixels where $R_{31}/\sigma_{R_{31}} \geq 3$, where $\sigma_{R_{31}}$ includes the RMS uncertainties for both the CO(1-0) and CO(3-2) flux measurements as well as the absolute flux calibration uncertainties. The map of the core region obtained with naturally weighted intensity maps is shown overlaid on the lower resolution map obtained with 1\farcs 5 {\it uv}-tapered intensity maps. The beams sizes for the two maps are shown in the lower right. In both cases, CO (1-0) and CO (3-2) intensity images were obtained with a matching beam size. Black contours trace the UV emission seen in the \textit{HST} F336W image. White contours trace the Band 6 continuum emission from the AGN point source. 
}
\end{figure*}

In order to study the CO line ratios in MACS 1931 in more detail, in Figure \ref{fig:R31_Map} we map $R_{31}$ in all pixels where the value of $R_{31}/\sigma_{R_{31}} \geq 3$, after accounting for both RMS uncertainties in the CO(1-0) and CO(3-2) images and the uncertainty in the absolute flux calibration. To examine $R_{31}$ in as much of the BCG as possible, we overlaid the ratio map obtained with beam-matched, naturally weighted images onto the ratio map obtained with beam-matched, 1\farcs{5} {\it uv}-tapered images. We superimposed the contours from the \textit{HST} F336W UV image on the $R_{31}$ map. 

In the BCG core, there are several notable bright UV knots to the north of the AGN and a depression in UV output to the south, which we attribute to a dust lane. $R_{31}$ loosely traces both sets of features, suggesting local enhancement of the CO lines ratios by star formation to the north and by the process involved in creating the dust lane to the south. $R_{31}$ is lower in the tail than in the core, and roughly drops off with increasing distance from the AGN point source. However, the tail and regions to the south and east of the core show areas of enhancement in $R_{31}$. 

Spatial variation in the excitation of CO in MACS 1931 has some similarities to the variation observed by \cite{2017Lim_Perseus} in NGC 1275, who found that $R_{32}$ varied between $\sim 1$ near the center of the galaxy and $\sim 0.5$ further out. Furthermore, $R_{21}$ values observed in NGC 1275 by \cite{2008Salome_NGC1275} varied between 0.6 and 1.7 in the center, but were observed to be as high as 2.4 at offset positions. MACS 1931 similarly shows enhancements of $R_{31}$ at positions several 10s of kpc from the BCG center.

There are likely varying contributions to the molecular gas excitation in cool-core BCGs from star formation, from the AGN, and from collisional excitation due to AGN-feedback moderated turbulence (e.g., \cite{2017Lim_Perseus}). This variation dominates the errorbar in the average $R_{31}$ value for the BCG subsample (green data point) shown in Figure \ref{fig:CO_SLED}. 

Molecular gas excitation is also a potentially useful way to compare BCGs in cool-core clusters like MACS 1931 to BCG progenitors in high-redshift protoclusters. Protoclusters can contain galaxies with even more massive molecular gas reservoirs than that found in MACS 1931, but the origins of the molecular gas and how it interacts with the cluster environment may be substantially different \citep{2016Emonts_Spiderweb, 2017Dannerbauer_HighZGas}. For example, compared to MACS 1931, the molecular gas feature in the central radio galaxy of a $z = 2.2$ protocluster known as the Spiderweb Galaxy is significantly more excited, and contains ten times as much molecular gas ($2\pm0.2\times10^{11}$ M$_{\odot}$) \citep{2018Emonts_Spiderweb}. $R_{41}$ in the Spiderweb is $\sim 1$, suggesting more extreme molecular gas excitation conditions than in MACS 1931. However, among the studied high redshift radio galaxies (a population which is also thought to contain progenitors of low-redshift BCGs), a wide range of molecular gas masses ($\sim 10^{10}-10^{11}$ M$_{\odot}$) have been reported, along with a wide range of excitation conditions \citep{2008Miley_HzRGs}. The significant differences in both the amount of gas and excitation of the gas suggests that the processes involved in producing star formation and molecular gas may differ in cool-core clusters with quasar-like AGN and in protoclusters.

\subsection{What Are the Origins of BCG Dust, and What Are Its Effects on the Molecular Gas?}

Observations of dust continuum emission in Bands 6 and 7 reveal that BCG dust is concentrated in regions that also contain multiphase gas and star formation. The dust-to-molecular gas ratio in the MACS 1931 BCG is $\sim 0.01-0.1$, based on the total molecular gas mass reported here and the dust mass derived in \cite{2017Fogarty_CLASH}. A similar dust-to-gas ratio in the range $\sim 0.04-0.1$ is observed in the Phoenix cluster BCG (\cite{2016Russell_Phoenix, 2017Fogarty_CLASH}). However, the relative distributions of dust and gas in MACS 1931 does not appear to be uniform, since the relative FIR emission in Bands 6 and 7 due to dust along the H$\alpha$ tail is much greater than that in seen in the core (excluding possible contributions to dust emission from the AGN point source that were subtracted out).

\begin{figure*}
\begin{center}
\begin{tabular}{c}
\includegraphics[height=7cm]{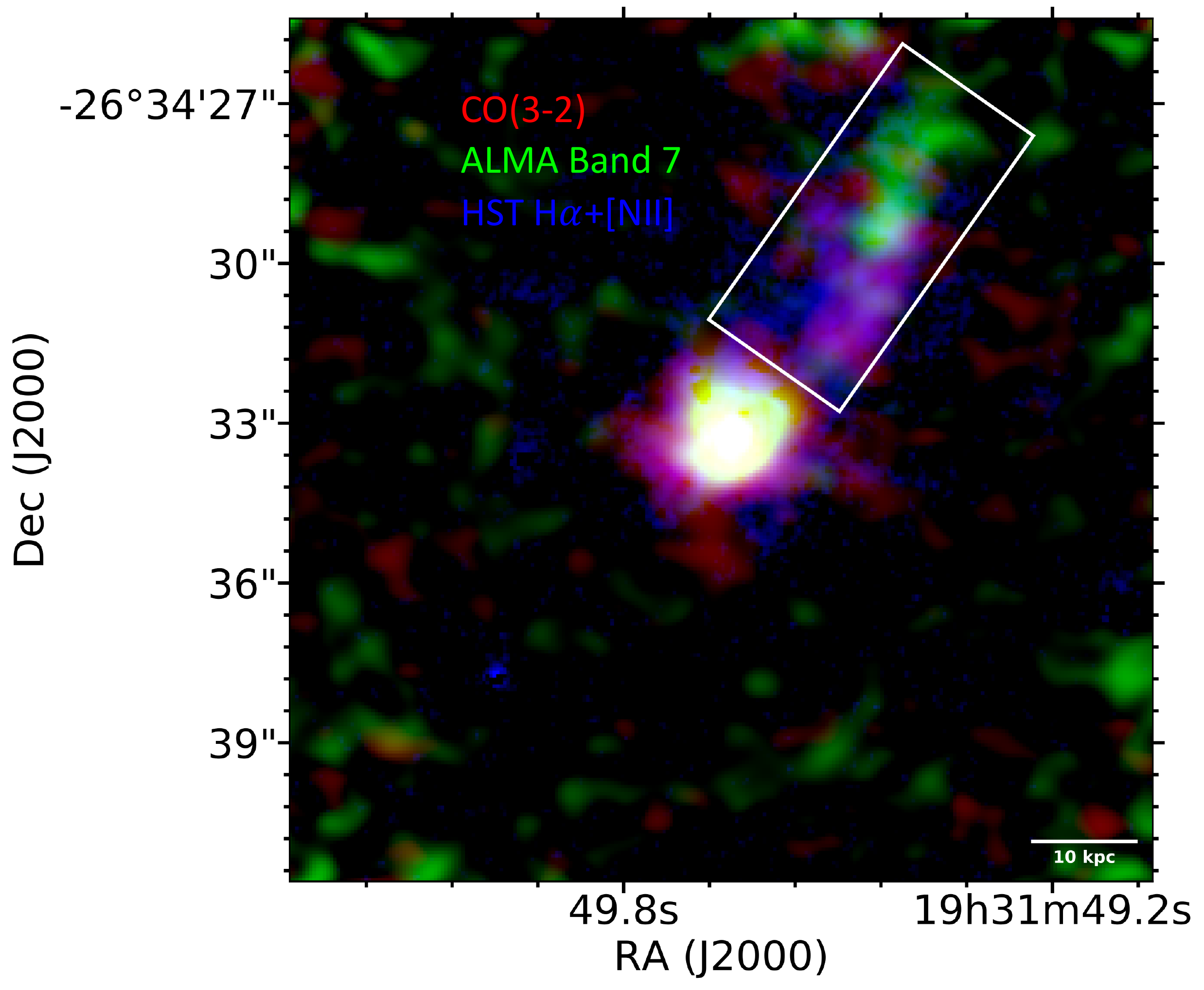}
\includegraphics[height=7cm]{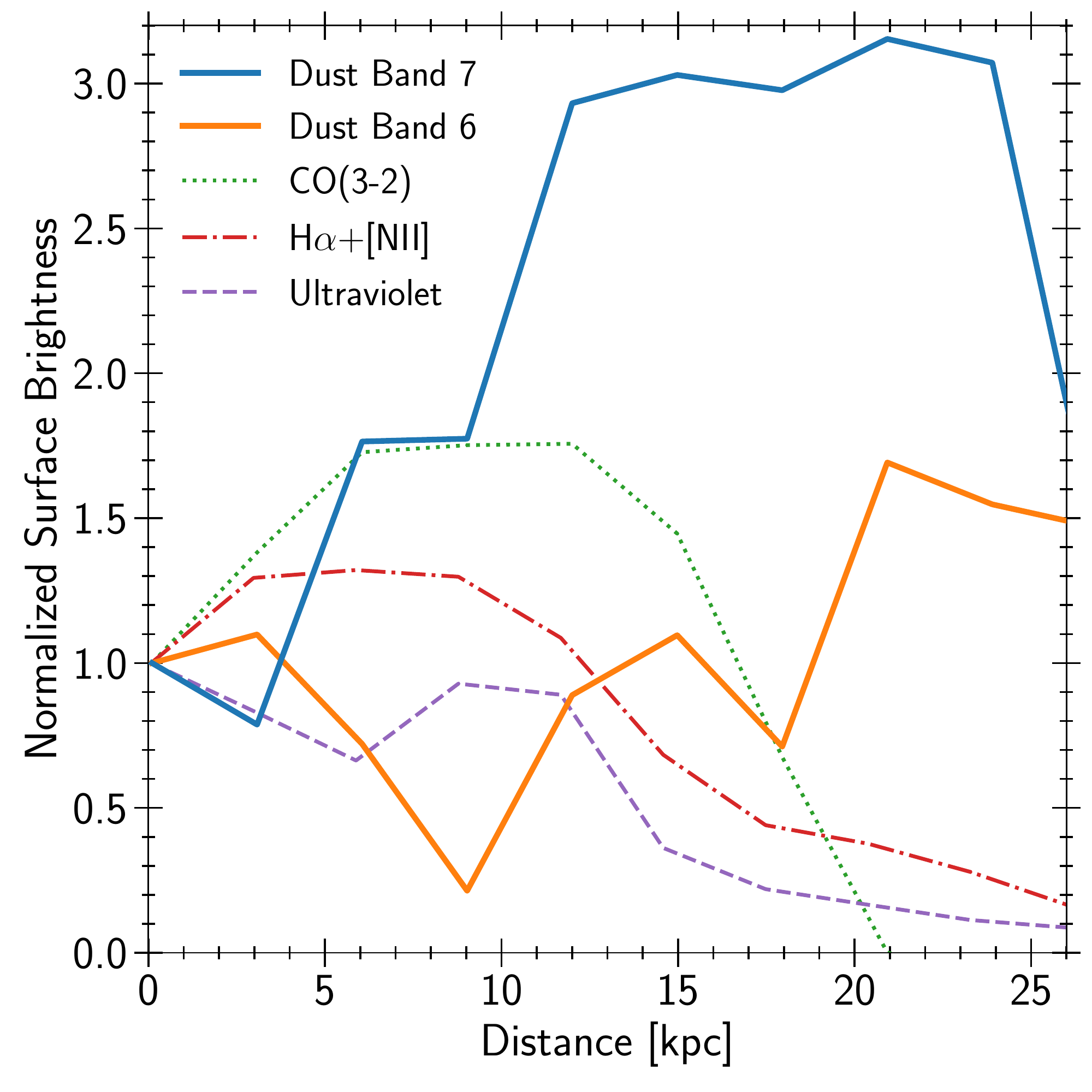}
\end{tabular}
\end{center}
\caption[]
{ \label{fig:Tail_SB} \textit{Left:} RGB composite of the CO(3-2) intensity (red), Band 7 extended continuum flux (green) and \textit{HST} broadband-subtracted H$\alpha$+[\ion{N}{2}] (blue). The white box shows the region used to extract the surface brightness profiles shown in the right-hand plot. \textit{Right:} Surface brightness profiles of the ALMA Band 6 and 7 extended continuum fluxes are shown compared to the profile of CO(3-2), H$\alpha$+[\ion{N}{2}], and \textit{HST} F336W UV. Profiles are normalized to the value at the edge of the rectangular slice nearest to the BCG core, and are binned in 0\farcs6 (3 kpc) intervals in order to smooth over variations due to noise and small structures.}
\end{figure*}

There is substantial variation in the relative FIR flux due to dust compared to CO, UV, and H$\alpha$ flux in the H$\alpha$ tail extending northwest of the BCG. In Figure \ref{fig:Tail_SB}, we plot the normalized FIR, CO(3-2), UV, and H$\alpha$+\ion{N}{2} surface brightness profiles in a $3''$ wide rectangular slice along the H$\alpha$ tail. Dust emission is elevated in the $\sim 10$ kpc of the tail furthest from the core compared to the UV, H$\alpha$+\ion{N}{2} and CO(3-2) surface brightness profiles. This result suggests that the dust concentration is enhanced in the outskirts of the tail.

Dust in the BCG either forms \textit{in situ}, is launched out of the BCG core by jets, or is the result of a combination of processes. \textit{In situ} cooling requires that dusty winds from AGB stars are preserved long enough to end up in the surrounding dusty nebular structure \citep{2019Li_AGBDust}. This source of dust may be supplemented by supernova production boosted by the starburst \citep{2003Dunne_SNDust}. Alternatively, dust in the BCG can arise from denser concentrations of dusty material dredged out of the center of the BCG by AGN jets \citep{1994Fabian_CFDust}. In this scenario, dusty material seeds the low-entropy gas uplifted by jets, which then induces condensation of the dust out of the resulting molecular clouds \citep{2011Voit_BCGMassLoss, 2014Voit_Feedback, 1994Fabian_CFDust}. The observed distribution of the dust in MACS 1931 relative to the atomic and molecular gas and UV emission from young stars suggests some uplift-dominated process is taking place, with dust emission most prominent in the core and at the furthest extremity of the tail, which is presumably the most recently uplifted body of material. In all likelihood, these processes co-mingle, with seed grains supplied by both jets and the AGB  population stimulating further condensation of dust out of the molecular gas.

Like the origin of dust, the persistence of dust in cool-core BCGs is a topic of debate, since sputtering induced by interactions with the hot ICM is expected to destroy dust on timescales of $\lesssim 10^{6}$ yr \citep{2011Voit_BCGMassLoss}. However, the very cold $\sim 10$ K temperatures that we derive for some of the dust in MACS 1931 suggests that a fraction of the dust in the BCG must be protected from sputtering or is in pockets of the ICM that are substantially colder than the $4.78\pm0.64$ keV temperature of the ICM plasma around the multiphase tail observed in \cite{2011Ehlert_MACS1931}. 

\cite{2004Montier_Dust} studied collisional heating of dust by the ICM, and concluded that for a wide range of ICM densities and temperatures above $\sim 1.7$ keV, the equilibrium temperature for dust interacting with the ICM does not depend substantially on grain size and is a function ICM density, $n_{e}$. Assuming that $n_{e} > 0.1$ cm$^{-3}$ (corresponding to the radial profile bin between $\sim 10$ and $\sim 30$ kpc in \cite{2011Ehlert_MACS1931}), the relation \cite{2004Montier_Dust} derives implies the dust equilibrium temperature must be at least $38$ K. A similar analysis by \cite{1990Dwek_ClusterDust} provides the same conclusion.

Meanwhile, \cite{2000Popescu_Dust} found that for small grain sizes the dust temperature distribution may peak at $\sim 10$ K, but they studied the core of the Virgo Cluster, which has an ICM temperature of $\sim 1.1$\,keV. In MACS 1931, a plausible scenario may thus be that the $\sim 10$ K dust in the ICM is either interacting with local volumes of $\lesssim 1$ keV material and has a large enough proportion of small grains that this population determines the Band 6 and 7 fluxes, or this dust is somehow protected from interacting with the hot ICM.

In the former case, we may hypothesize that the cold dust itself is a potential coolant. While the role of dust in promoting ICM cooling is still a subject of debate, recent findings suggest dust is capable of enhancing ICM cooling by up to $50\%$ \citep{2018Vogelsberger_Dust}. The multiphase tail to the northwest of the MACS 1931 BCG is located on a prominence in both ICM metallicity and X-ray flux, and the ICM surrounding the tail has the lowest entropy  (13-16 keV cm$^{2}$) of anywhere in the cluster \citep{2011Ehlert_MACS1931}. The extremely low entropy in the ICM immediately surrounding the tail (see Figure 5d in \cite{2011Ehlert_MACS1931}) suggests the potential for further condensation, although it may also imply a conductive or turbulent mixing layer. In the event that the very cold dust in the multiphase tail is tracing pockets of ICM that are several keV colder than this already low-entropy gas,
it is possible that the dust itself is promoting condensation of the ICM onto the multiphase tail, and the growth of dust out of the molecular gas is sufficient to allow the extended dust to persist.

\begin{figure}
\begin{center}
\begin{tabular}{c}
\includegraphics[width=3.4in]{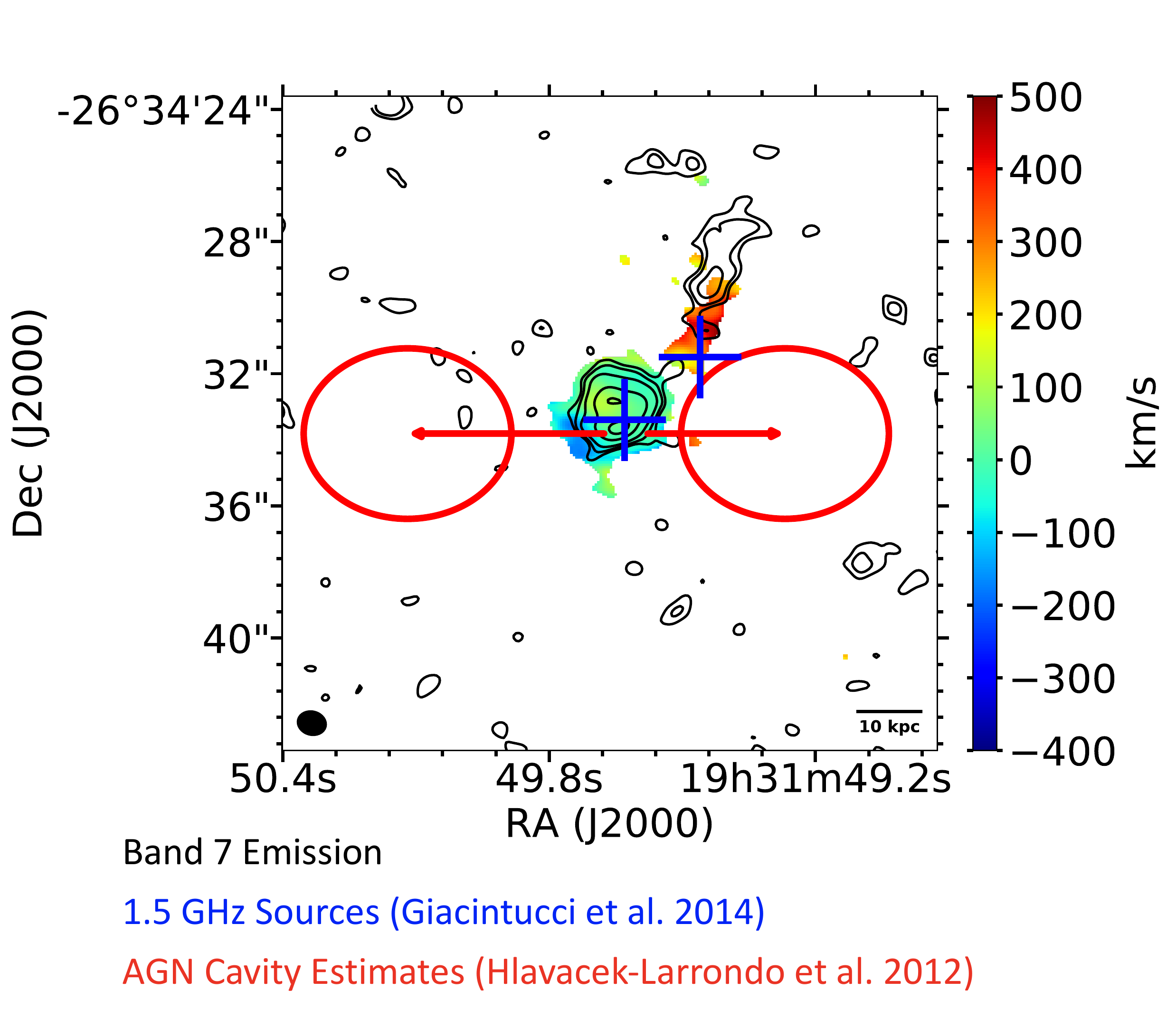}
\end{tabular}
\end{center}
\caption[]
{ \label{fig:Discussion_Figure} The CO(3-2) velocity field with dust emission overlaid along with indicators for features visible in X-ray and 1.5 GHz emission. Contours showing the Band 7 continuum emission due to extended dust are shown in black, at the 2, 3, 5, 10, and 15$\sigma$ levels. Blue crosses indicate the positions of compact 1.5 GHz emission (\cite{2014Giacintucci_Minihalos} and \textit{private communication}). Red lines connect the X-ray brightness peak to the positions of the X-ray cavity centers and red ellipses outline the extent of the X-ray cavities based on estimates of the radial separations of the cavities from the center of the cluster and cavity semi-major and semi-minor axes in \cite{2012HL_Cavities}. Extended molecular gas and dust filaments in this BCG are located nearly 90$^{\circ}$ from the axis between the X-ray cavities extending from the AGN.} 
\end{figure}

\subsection{Correlation Between the Dusty Multiphase Tail and Features in the Radio Mini-Halo}

As previously observed in \cite{2014Giacintucci_Minihalos} and \cite{2018Yu_JVLA}, there is a radio mini-halo in the core of MACS 1931 which may be associated with the feedback processes occurring in this system. Radio mini-halos are small, typically diffuse, structures around cool-core BCGs that extend across some or all of the cluster core \citep{2016Gitti_Minihalos}. While radio mini-halos are thought to occur in most cool-core cluster cores \citep{2017Giacintucci_Minihalos}, the mini-halo in MACS 1931 is unusual for having two compact 1.5 GHz sources \citep{2014Giacintucci_Minihalos}: an 11.6 mJy source on the position of the BCG core and a 2.5 mJy source on the position of the tail as shown Figure~\ref{fig:Discussion_Figure}. We note that the 1.5 GHz flux predicted by the \cite{2006Schmitt_RadioSFR} relation for a SFR of 250 M$_{\odot}$ yr$^{-1}$ is $\sim 1$ mJy, so neither component can be easily explained solely by star formation in the BCG. 

While the origins of radio minihalos in cool-core clusters are not well known, observations and simulations suggest that these objects are due to turbulent re-acceleration of relativistic electrons in the ICM, which arise from the relativistic plasma injected into the system by AGN jets \citep{2007Fujita_Minihalos, 2013ZuHone_Minihalos, 2019vanWeeren_DiffuseRadio}. The presence of the smaller compact 1.5 GHz component is striking because of its correspondence with the multiphase tail. It is likely that this feature arises out of interactions between the relativistic electron population of the ICM and one or more components of the multiphase tail, although it is not clear what is driving the acceleration of the electron population. We point out this correlation because it is intriguing, and merits further investigation.

\section{Conclusions}

Our new ALMA Band 3, 6, and 7 observations of the molecular gas and dust in the MACS 1931 BCG reveal a large $1.9 \pm 0.3 \times 10^{10}$ M$_{\odot}$ reservoir of molecular gas found principally in the BCG core and in an elongated tail stretching $\sim 30$ kpc to the northwest. Continuum images in Bands 6 and 7 reveals large concentrations of dust tracing the ICM, and particularly cold ($\sim 10$ K) dust in portions of the tail. 

Furthermore, dust and molecular gas features in MACS 1931 broadly correlate with the positions of UV knots and H$\alpha$ filaments seen in the \textit{HST} images obtained by the CLASH program, demonstrating that the material in the core of this cluster is both multiphase and forming stars. 

The principal findings of our analysis of our ALMA data are:
\begin{itemize}
    \item The CO flux in the MACS 1931 BCG is strongly detected in all three bands observed. The CO(1-0) flux is $3.45\pm 0.47$ Jy\,km\,s$^{-1}$, the CO(3-2) flux is $29.0\pm 2.8$ Jy\,km\,s$^{-1}$, and the CO(4-3) flux is $33.7\pm3.5$ Jy\,km\,s$^{-1}$ (see Table \ref{table:Line_Params}).
    \item Molecular gas in the BCG core is concentrated in knot-like structures that have velocities (relative to the BCG optical redshift) varying between $-98\pm 11$ km\,s$^{-1}$ and $258\pm 18$ km\,s$^{-1}$, which suggests these knots may be either chaotically infalling or orbiting. Molecular gas in the H$\alpha$ tail is likely falling inward at $\sim 300$ km\,s$^{-1}$.
    \item The typical velocity dispersions of CO(1-0) and CO(3-2) is $\sim 200$ km\,s$^{-1}$ After accounting for blurring induced by the synthesized beam, we find this result consistent with simulated cool-core BCGs. Our measured velocity dispersions are likely dominated by the random motions of clouds that cannot be separated by our beam size.
    \item We measure CO line ratios in the central region of MACS 1931 of $R_{31} = 0.93\pm0.16$ and $R_{41} = 0.61\pm0.10$. $R_{31}$ is $\sim 1$ throughout much of the core and shows enhancements to $\sim 1$ in the tail (see Figure \ref{fig:R31_Map}).
    \item Extended rest-frame 892 $\mu$m and 640 $\mu$m continuum emission in Bands 6 and 7 imply the presence of dust in the H$\alpha$ tail. By examining the Band 7 to Band 6 flux ratios, we infer the presence of very cold, $\sim 10$ K dust in parts of the tail. We also find that dust emission is highest relative to CO emission at the end of the tail furthest from the BCG core.
\end{itemize}

Our analysis of the CO spectral line energy distribution reveals evidence for multiple gas excitation mechanisms in MACS 1931. Similar to QSOs and extreme ULIRGs, the CO(3-2) transition is highly excited \citep{2012Papadopoulos_ULIRG, 2015Rosenberg_ULIRG}, while the CO(4-3) is closer to what we expect from the star formation occurring in this system. We therefore suspect that there must be processes not related to the starburst contributing to the excitation of molecular gas in MACS 1931, although these processes may be less dominant than in some other ULIRGs. We study the spatial variation of $R_{31}$ and find that the molecular gas excitation levels also trace features seen in \textit{HST} UV photometry, where star formation processes excite CO. However, $R_{31}$ is also elevated in parts of the tail and to the south and east of the core, suggesting other excitation processes at work as well. Molecular gas excitation in MACS 1931 may be driven by a combination of processes, including star formation, AGN radiation, and interaction between the molecular gas and ICM.

We discuss possible options for the origins and current behavior of the molecular gas and dust in this system. The molecular gas depletion time in MACS 1931 is $\sim 80$ Myr, which is similar to the lifetime of its most recent starburst, suggesting that the period of enhanced feedback and cold gas condensation in this cluster has progressed in time at least to its midpoint. Furthermore, the infalling multiphase tail is not aligned with the jet axis implied by features consistent with east-west oriented X-ray cavities; morphological evidence suggests the gas may have originally formed at the interface between jet-inflated cavities and the ICM, similar to the scenario explored by e.g. \cite{2014McNamara_A1835, 2016Russell_Phoenix}, but must have migrated away from these interfaces, or formed long enough ago for the X-ray cavities it formed around to have dissipated.

The dust at the outer edge of the multiphase tail appears to have the highest concentration relative to CO. The location of an elevated dust concentration here, coupled with our inference that the gas in the tail is infalling, lead us to suspect that either the dust reservoir is growing here, or this material was the most recently uplifted from the core. Meanwhile, the temperature constraints of $T_d \sim 10$ K we place on the dust in certain regions allows us to conclude that at least part of the population of dust in MACS 1931 is either shielded from sputtering or is in regions of cooler, and possibly condensing, ICM plasma. Based on the analysis of \cite{2004Montier_Dust}, we conclude this dust is too cold to be interacting with ICM plasma hotter than $\sim 1.7$ keV, so if it is interacting with the ICM, it must be enveloped in volumes of gas that are at least 3 keV cooler than the mean ICM temperature around the tail \citep{2011Ehlert_MACS1931}.

Finally, we discuss the intriguing presence of a compact 1.5 GHz emission feature in the multiphase tail. This source cannot be explained by the levels of star formation present in MACS 1931, and its spatial coincidence with the tail implies possible interactions between the tail and electron population in the ICM that ought to be studied further.

Our ALMA observations provide critical insights into the nature of the molecular gas and dust in the center of a non-local cool-core galaxy cluster. By studying CO(1-0), CO(3-2), and CO(4-3), we are able to place constraints on the energy sources driving excitation of the molecular gas, and we are able to study the dynamics of molecular gas in a rare extreme BCG starburst. Our observations reveal the presence of cold dust in highly extended structures. Since these cold dust structures must either reside in volumes of ICM plasma that are substantially colder than the ambient ICM or imply shielding of the dust against sputtering, further study of cold dust in cool-core BCGs will be valuable for fully understanding cooling in clusters.

\vskip 12pt

The authors thank Piero Rosati and Miguel Verdugo for graciously providing the VLT/MUSE spectrum of the MACS 1931 BCG to us in advance of its publication.
This paper makes use of the following ALMA datasets: ADS/JAO.ALMA\#2016.1.00784.S and ADS/JAO.ALMA\#2017.1.01205.S. ALMA is a partnership of ESO (representing its member states), NSF (USA) and NINS (Japan), together with NRC (Canada), MOST and ASIAA (Taiwan), and KASI (Republic of Korea), in cooperation with the Republic of Chile. The Joint ALMA Observatory is operated by ESO, AUI/NRAO and NAOJ.
This research was supported, in part, by NASA grant HSTGO-12065.01-A. The CLASH Multi-Cycle Treasury Program is based on observations made with the NASA/ESA \textit{Hubble Space Telescope} and which is operated by the Space Telescope Science Institute. 
K.F. acknowledges financial support from the Troesh Prize Postdoctoral Fellowship from the California Institute of Technology.
H.D. acknowledges financial support from the Spanish Ministry of Economy and Competitiveness (MINECO) under the 2014 Ramon y Cajal program MINECO RYC-2014-15686.
This research made use of APLpy, an open-source plotting package for Python \citep{2012Robitaille_aplpy}.
The authors would like to thank the anonymous referee for their quick response to our submitted manuscript.

\bibliography{MACS1931_ALMA_Refs}

\end{document}